\documentclass[reprint, amsmath, amssymb, aps, pre]{revtex4-2}

\usepackage{graphicx}
\usepackage{dcolumn}
\usepackage{bm}
\usepackage{xcolor}
\usepackage{amsmath}
\usepackage{soul}
\DeclareMathOperator\erfc{erfc}

\begin{document}

\title{Direct Evaluation of the Phase Diagrams of Dense Multicomponent Plasmas by Integration of the Clapeyron Equations}
\author{Simon Blouin}
\author{J{\'e}r\^{o}me Daligault}
\affiliation{Los Alamos National Laboratory, PO Box 1663, Los Alamos, NM 87545, USA}

\date{\today}

\begin{abstract}
Accurate phase diagrams of multicomponent plasmas are required for the modeling of dense stellar plasmas, such as those found in the cores of white dwarf stars and the crusts of neutron stars. Those phase diagrams have been computed using a variety of standard techniques, which suffer from physical and computational limitations. Here, we present an efficient and accurate method that overcomes the drawbacks of previously used approaches. In particular, finite-size effects are avoided as each phase is calculated separately; the plasma electrons and volume changes are explicitly taken into account; and arbitrary analytic fits to simulation data as well as particle insertions are avoided. Furthermore, no simulations at ``uninteresting'' state conditions, i.e., away from the phase coexistence curves, are required, which improves the efficiency of the technique. The method consists of an adaptation of the so-called Gibbs--Duhem integration approach to electron--ion plasmas, where the coexistence curve is determined by direct numerical integration of its underlying Clapeyron equation. The thermodynamics properties of the coexisting phases are evaluated separately using Monte Carlo simulations in the isobaric semi-grand canonical ensemble (NPT$\Delta \mu$). We describe this Monte Carlo-based Clapeyron integration method, including its basic physical and numerical principles, our extension to electron--ion plasmas, and our numerical implementation. We illustrate its applicability and benefits with the calculation of the melting curve of dense carbon/oxygen plasmas under conditions relevant for the cores of white dwarf stars and provide analytic fits to implement this new melting curve in white dwarf models. While this work focuses on the liquid--solid phase boundary of dense two-component plasmas, a wider range of physical systems and phase boundaries are within the scope of the Clapeyron integration method, which had until now only been applied to simple model systems of neutral particles.
\end{abstract}

\maketitle

\section{Introduction}
When a plasma comprised of different ionic species crystallizes, the newly-formed solid generally has a different composition than the coexisting liquid. This composition change is characterized by a phase diagram. Accurate phase diagrams of multicomponent plasmas are essential for the study of dense astrophysical objects, and in particular for white dwarf stars. The crystallization of their C/O interiors leads to the formation of a solid core enriched in O \citep{stevenson1980,mochkovitch1983,tremblay2019}. The separation of the C and O ions releases gravitational energy that delays the cooling of white dwarf stars by $\simeq 1\,{\rm Gyr}$ \citep{garcia1988b,segretain1994,isern1997,isern2000,althaus2012,blouin2020}, with important implications for the use of white dwarfs as cosmic clocks \citep{garcia1988,fontaine2001}. A similar phase separation process is expected to occur for other minor chemical species in white dwarf interiors (chiefly $^{22}$Ne and $^{56}$Fe) \citep{isern1991,xu1992,segretain1996}. Phase diagrams of dense multicomponent plasmas are also needed for the study of accreting neutron stars \citep{horowitz2007}.

Over the few last decades, several methods have been proposed and used to map the phase diagrams of dense astrophysical plasmas. They can roughly be classified into three categories: density-functional methods \citep{barrat1988,segretain1993}, Monte Carlo-based (MC) techniques \citep{ichimaru1988,ogata1993,dewitt1996,dewitt2003,medin2010}, and molecular dynamics (MD) approaches \citep{horowitz2007,horowitz2010,hughto2012,schneider2012}. Density-functional techniques, which rely on analytical models of the free energies of the relevant phases are inherently more approximate than MC and MD simulation techniques. MC-based methods generally consist of constructing analytic fits to results from MC simulations in the canonical (NVT) ensemble in order to obtain an analytical model for the Helmholtz free energies of the coexisting phases. The free energies of the liquid and solid are then compared to identify the location of the coexistence curve. A major limitation of this approach is that the resulting phase diagram is sensitive to the (somewhat arbitrary) choices of analytic functions used to interpolate the MC data. This can even affect the qualitative shape of the phase diagram. For example, due to minute differences in their interpolation functions for the internal energies of binary ionic mixtures (BIMs), ref.~\cite{ogata1993} concludes that the C/O phase diagram is of the azeotrope shape, while ref.~\cite{dewitt1996} finds that it is spindle shaped. This extreme sensitivity is due to the very small differences between the free energies of the liquid and solid phases near the coexistence conditions, and highlights the need for accurate ``at-parameter'' calculations.

Two-phase MD methods such as those used by the Horowitz \textit{et al.} group \citep{horowitz2007,horowitz2010,hughto2012,schneider2012} have the advantage of not requiring any interpolation. Typically, a liquid and a solid phase are initially put in contact and then evolved in time in the microcanonical or canonical ensemble. The particles diffuse through the liquid--solid interface and eventually a state of equilibrium is reached, which allows to pinpoint the coexistence conditions that characterize the liquid--solid transition. However, a practical drawback of MD approaches is their steep computational cost. Because of this, the phase diagram can only be partially sampled, leading to rather coarse coexistence curves (e.g., see Fig.~2 of ref.~\cite{horowitz2010}). This can be a problem for astrophysical applications. For instance, the coarse sampling of the melting curves of the C/O phase diagram of ref.~\cite{horowitz2010} leads to sizeable uncertainties on the gravitational energy released by the O sedimentation process in white dwarfs. Similarly, a fine sampling of phase diagrams is required to precisely identify the location of interesting features such as an azeotropic or eutectic point. In addition, MD simulations with liquid--solid interfaces are subject to detrimental finite-size effects. While these artifacts can in principle be mitigated using a large enough number of particles, the required number of particles and the cost of the corresponding simulations often prohibit detailed studies of this type \cite{horowitz2010}.

Previous approaches (both MC- and MD-based) commonly assumed a constant volume during the phase transition. This simplifies the problem: the electronic background does not need to be treated explicitly as the electronic density remains constant. Only the screening effect of the electrons on the bare ion--ion interactions was usually included (using a Yukawa potential instead of a Coulomb potential). But this simplification is not strictly correct. Phase transitions occur at constant pressure and are accompanied by volume changes. That being said, in the particular case of dense astrophysical plasmas, where the total pressure is dominated by the degenerate electron gas, the constant volume approximation is well justified (we demonstrate this point explicitly in Appendix~\ref{sec:validation}, see also refs.~\cite{ogata1993,medin2010}).

An alternative technique to calculate phase diagrams is the so-called Gibbs--Duhem integration method (we prefer the term ``Clapeyron integration method''), where the coexistence curve is obtained by direct integration of the appropriate Clapeyron relation \cite{kofke1993a,kofke1993b}. As discussed below, this new approach to calculate the phase diagrams of dense plasmas is largely free of the limitations that characterize the competing methods outlined above.  While the method has so far been applied with great success to simple models of neutral mixtures \citep{agrawal1995a,agrawal1995b,agrawal1995c,hitchcock1999,
lamm2001,lamm2001b,lamm2002,lamm2004}, it has never been used for electron--ion plasmas. Adapting this method to charged systems is not as straightforward as substituting an interaction potential by another. In particular, electrons must be explicitly included in the calculations, since the method involves volume changes and ionic identity changes that affect the electronic background. This added complexity has its advantages (it is physically more satisfying to perform all calculations at constant pressure rather than at constant volume), but requires additional care. 

The central goal of this paper is to explain how the Clapeyron integration technique can be adapted to map the phase diagrams of dense plasmas. This work is a companion paper to ref.~\cite{blouin2020} (where we presented the astrophysical implications of our new C/O phase diagram) that provides a detailed account of the method. Because the Clapeyron integration approach is not commonly used, we begin in Section~\ref{sec:gd} with a self-contained and pedagogical introduction to this technique instead of simply referring the reader to the original papers; for clarity and completeness, a number of technical details are given in the appendices. The application to electron--ion plasma mixtures requires some care and the needed adaptations are highlighted. Section~\ref{sec:gd} also includes an illustration of the method and its inner workings using a simple analytic model of plasma mixtures. After this general discussion, we delve into the specifics of the plasma model that we use to compute the phase diagrams of dense plasmas (Section~\ref{sec:theory}, with additional details in Appendices~\ref{sec:partfunc} and \ref{sec:yukawa}). We then describe the MC method that we have implemented for this purpose (Section~\ref{sec:mc_code}). Extensive tests of our code are presented in Appendix~\ref{sec:validation}. As an example application of this new simulation capability, we present the calculation of the phase diagram of the C/O interior of white dwarf stars in Section~\ref{sec:CO}, where we also provide useful analytic fits for implementation in white dwarf models. Finally, a short summary is given in Section~\ref{sec:conclusion}.

\section{Clapeyron Equation Integration Method}
\label{sec:gd}

\subsection{Qualitative Overview of the Method}
A Clapeyron equation is a relation between the intensive thermodynamic variables that characterize the conditions of coexistence between two or more thermodynamic phases of a physical system. The present method consists in numerically integrating this Clapeyron equation along the coexistence curve. At each equilibrium point along the coexistence curve, the thermodynamic properties of each coexisting phase are calculated simultaneously, but separately (we perform this step using MC simulations, see Section~\ref{sec:mc_code}). This allows to numerically evaluate and integrate the Clapeyron equation from one state point on the coexistence curve to a neighboring point on the curve. Pairs of MC simulations for the liquid and solid phases are computed in succession until the coexistence line is fully mapped.

The thermodynamic properties of the system only have to be evaluated at the coexistence conditions, meaning that no uninteresting state points are calculated. This has three major advantages compared to the above-mentioned standard methods where free energy models are built by interpolating between many intermediate state points: (1) it is more efficient from a computational point of view (less states to simulate), (2) no arbitrary interpolation is required, thereby increasing the numerical accuracy of the calculation, and (3) all thermodynamic properties of the system at the phase transition are readily available at no additional cost. 

The first two advantages given above are also shared with the two-phase MD approach. However, the Clapeyron integration approach is also free of what is probably the greatest limitation of the two-phase MD technique. Since each phase is treated independently (at each coexistence point, one MC simulation is performed for the liquid phase and another one for the solid phase), there is no liquid--solid interface to simulate. This eliminates a major contributor to detrimental finite-size effects. Finite-size effects can be easily mitigated in (isotropic) single-phase simulations. Note also that the MC calculations needed to integrate the Clapeyron equation are relatively cheap, which allows a finer sampling of the phase diagram than interfacial MD simulations and a better resolution of its interesting features (e.g., an azeotropic point).

The MC simulations needed to integrate the Clapeyron equation are performed in an isomolar ensemble: no particle insertions or deletions are needed. This constitutes another important advantage of this approach, as methods that require transfers of particles are not practical for strongly interacting systems such as those in which we are interested \cite{panagiotopoulos1994}.

Finally, all calculations in the Clapeyron integration approach are performed at constant pressure. This is to be contrasted with most phase transition calculations where a constant volume is assumed, while in reality phase transitions virtually always occur at constant pressure and imply volume changes. Even if the constant volume approximation is often very accurate, it is inherently more satisfying to perform all calculations in the correct thermodynamic ensemble and it makes the method applicable to a broader range of systems.

\subsection{Thermodynamics of electron--ion plasmas} \label{Sec_II_B}
As we shall see, the Clapeyron integration method is formulated in an isobaric semi-grand statistical ensemble (NPT$\Delta \mu$), i.e., at constant pressure $P$, constant temperature $T$, constant total number of particles $N$, and constant relative chemical potentials $\Delta\mu_a$. In this ensemble, the volume of the system $V$ and the number of particles $N_a$ of the different particle species fluctuate. Therefore, the application of the method to an electron--ion plasma raises questions regarding the inclusion of electrons in the calculation. Both the allowed variations in volume and in particle numbers imply variations in the electronic density. One consequence of those variations is that the screening length used to screen the ionic interactions can no longer be assumed to be constant as in standard methods. While this is obvious in the case of volume fluctuations, the effect of fluctuations of particle numbers is more subtle. For the finite-size calculations to be physically meaningful and have a well-defined thermodynamic limit (i.e., $\{N_a\},V\to \infty$ at constant density $\{N_a\}/V$), it is necessary to enforce the global neutrality of the system. In other words, the thermodynamic limit should be taken at constant $N_Z=\sum_a{Z_aN_a}=0$, where the sum includes the plasma electrons and $Z_a e$ is the charge of species $a$. For our purpose, we found it useful to constrain the number of electrons. If $\left( \{ N_i \} , \{ Z_i \} \right)_{i=1,\dots,c}$ denotes a given ionic composition of the plasma, with $c$ the number of ionic species, $N_i$ the fluctuating number of ions of species $i$ and $Z_i e$ the charge of each species, then we enforce the number of electrons $N_e = \sum_{i=1}^c Z_i N_i$ to ensure neutrality.

With this choice, the independent extensive variables are the entropy $S$, the volume $V$, and the number of ions of each species $\{ N_i \}$. The internal energy is given by
\begin{equation}
U \left( S,V, \{ N_i \} \right) = TS - PV + \sum_{i=1}^c N_i \mu_i,
\end{equation}
where $\mu_i$ is the electrochemical potential of species $i$, defined as the sum of the ionic chemical potential and an electronic contribution,
\begin{equation}
\mu_i = \mu_{{\rm ion},i} + Z_i \mu_e.
\end{equation}
With these variables, the equilibrium conditions between a liquid $(\ell)$ and a solid $(s)$ phase are $P^{\ell} = P^s$, $T^{\ell} = T^s$, and $\mu^{\ell}_i = \mu^s_i$ for all $i=1,\dots,c$. Here, we restrict the discussion to the coexistence line between a liquid and a solid phase, although what follows applies to other phase boundaries as well. In addition, the case of systems of neutral particles \citep{kofke1993a,kofke1993b,agrawal1995a,agrawal1995b,agrawal1995c,hitchcock1999,
lamm2001,lamm2001b,lamm2002,lamm2004} is recovered by setting $\mu_e=0$ in the previous and following equations.

\subsection{Clapeyron Equation}

We now turn to the derivation of the Clapeyron equations that form the backbone of our integration technique. The latter are conveniently derived from the Gibbs--Duhem relation among the temperature $T$, pressure $P$, and chemical potentials $\mu_i$ \citep{denbigh1981}, and below we limit ourselves to examples relevant to our purpose. For a $c$-component mixture, the Gibbs--Duhem relation can be expressed as
\begin{equation}
d \mu_1 = -s dT + v dP - \sum_{i=2}^c x_i d \left(\mu_i - \mu_1 \right),
\label{eq:gd}
\end{equation}
where $s = S/N$ is the entropy per ion (with $N = \sum_{i=1}^c N_i$), $v = V/N = 1/n$ is the volume per particle, and $x_i = N_i/N$ is the number concentration of species $i$. For a one-component system, this relation directly leads to the usual form of the Clapeyron equation. For the solid and liquid phases to coexist, a change in temperature must cause a change in pressure such that the chemical potentials $\mu^{\ell}$ and $\mu^{s}$ of the liquid and solid phases remain equal. From Eq.~\eqref{eq:gd}, this implies $s^{\ell} dT - v^{\ell} dP = s^{s} dT - v^s dP$ along the coexistence line, and in turn
\begin{equation}
\frac{d P}{d T} = \frac{s^{\ell} - s^s }{v^{\ell}-v^s} = \frac{L_m}{T \left( v^{\ell}-v^s \right)},
\end{equation}
where $L_m$ is the latent heat released per particle.

For multicomponent systems, different Clapeyron equations between two field variables can be similarly derived by fixing the other field variables to their phase equilibrium values. For a two-component mixture, fixing $P$ leads to the Clapeyron relation
\begin{equation}
\left. \frac{dT}{d\left( \mu_2 - \mu_1 \right)} \right|_P = - \frac{x_2^{\ell} - x_2^s}{s^\ell - s^s},
\label{eq:clapeyron2}
\end{equation}
which describes the relation between $T$ and the difference in the chemical potentials of the two species, $\mu_2 - \mu_1$, along the coexistence line. Similarly, other Clapeyron relations can be derived for $c>2$ component mixtures. We detail the case of a three-component mixture in Appendix~\ref{sec:3component}.

In order to exploit Eq.~\eqref{eq:clapeyron2}, it will be beneficial to work with a thermodynamic potential that explicitly depends on $\mu_i - \mu_1$. This can be achieved using the isobaric semi-grand canonical potential,
\begin{multline}
{\cal{A}}(T,P,N,\{\tilde{\mu}_i-\tilde{\mu}_1\}_{i=2,\dots,c}) =\\ U-TS+PV-\sum_{i=2}^{c}{(\tilde{\mu}_i-\tilde{\mu}_1)N_i} = N \tilde{\mu}_1,
\label{eq:muref_first}
\end{multline}
where we have defined $\tilde{\mu}_i = \mu_i + \mu_i^{\rm ref}$. Here, $\mu_i^{\rm ref}$ is a given reference chemical potential that we have added to the formalism to deal with plasma systems. $\mu_i^{\rm ref}$ is innocuous at the level of the theory, but, as we shall see in Section~\ref{sec:mc_code}, it plays an important role in the numerical applications. In addition, it will be more convenient in practice to work in terms of fugacity fractions $\xi_i$ instead of chemical potentials. Let the fugacity of species $i$ be
\begin{equation}
f_i = e^{\beta \left( \tilde{\mu}_i - \mu_i^0 \right)},
\label{eq:fugacity_def}
\end{equation}
where $\beta = 1/\left( k_B T \right)$, $e^{- \beta \mu_i^0} = V/ \Lambda_i^3$, $\Lambda_i$ is the thermal de Broglie wavelength, $\Lambda_i  = h/\left(2 \pi M_i k_B T\right)^{1/2}$,
$k_B$ is Boltzmann's constant, $h$ is Planck's constant, and $M_i$ is the mass of species $i$. Then, in the general case of a $c$-component mixture, the fugacity fraction is defined as
\begin{equation}
\xi_i = \frac{f_i}{\sum_{i=1}^c f_i}.
\label{eq:fugacityfrac_def}
\end{equation}
Unlike the chemical potentials that generally can take any real values, the fugacity fractions are constrained to vary between $0$ and $1$ (i.e., $0 \leq \xi_i \leq 1$), which is useful numerically. Moreover, one can show that $\xi_i=0$ ($\xi_i=1$) when the number concentration $x_i=0$ ($x_i=1$).

Using these new definitions, the $c$-component Gibbs--Duhem equation [Eq.~\eqref{eq:gd}] can be written as
\begin{equation}
d\ln\left(\sum_{i=1}^c{f_i}\right)=h_r d\beta+\frac{\beta P}{n}d\ln P-\sum_{i=1}^{c}{x_i\frac{d\xi_i}{\xi_i}},
\label{eq:gd2}
\end{equation}
where $h_r=h-\sum_{i=1}^c{x_i\frac{d}{d\beta} \beta(\mu_i^0+\mu_i^{\rm ref})}$, with $h=(U+PV)/N$ the enthalpy per ion. When expressed in terms of the fugacity fractions, the two-component Clapeyron relation of Eq.~\eqref{eq:clapeyron2} reads \cite{hitchcock1999}
\begin{equation}
\left. \frac{d \beta}{d \xi_2} \right|_{P}=\frac{x_2^{\ell}-x_2^{s}}{ \xi_2(1-\xi_2)(h_r^{\ell}-h_r^{s})}.
\label{eq:clapeyron}
\end{equation}
For a given pressure, this form of the Clapeyron equation describes how the temperature changes with the fugacity fraction along the liquid--solid coexistence line. This is the equation that we will integrate to map the phase diagrams of two-component plasmas. The properties of the fugacity fractions imply that in order to map the phase diagram of a given two-component plasma, Eq.~\eqref{eq:clapeyron} simply needs to be integrated from $\xi_2=0$ to $\xi_2=1$.

To carry out the Clapeyron integration, the concentrations and enthalpies that appear in the right-hand side of Eq.~\eqref{eq:clapeyron} have to be evaluated for the fixed $P$, $T$, and $\xi_i$'s that characterize each state point along the coexistence curve. It is for this reason that it makes sense to work in the isobaric semi-grand canonical ensemble (NPT$\Delta \mu$). To link the microphysics of our system to the thermodynamic relations given above, we have \cite{briano1984}
\begin{equation}
{\cal{A}}=-k_BT\ln {\cal{Q}},
\end{equation}
with the partition function,
\begin{align}
&{\cal{Q}} \left(T,P,N,\{\tilde{\mu}_i-\tilde{\mu}_1\}_{i=2,\dots,c}\right) = \nonumber\\
&\quad \quad  \int_0^\infty \Bigg[  \frac{dV}{V_0} \sum_{i_1=1}^c \dots \sum_{i_N=1}^{c} \frac{\prod_{i=1}^c{N_i!}}{N!}  \label{eq:calQ} \\
&\quad \quad \quad \times e^{-\beta{\cal{F}}\left[T,V,\{N_i\}_{i=1,\dots,c}\right]-\beta PV+\beta\sum\limits_{j=1}^{N}(\tilde{\mu}_{i_j}-\tilde{\mu}_1)} \Bigg] , \nonumber
\end{align}
where 
\begin{equation}
{\cal{F}} = - k_B T \ln \cal{Z}
\label{eq:calF}
\end{equation}
is the Helmholtz free energy and $\cal{Z}$ is the usual canonical partition function. The statistical average of a thermodynamic quantity $B$ is given by the following equation,
\begin{align}
&\langle B \rangle
=\frac{1}{\cal{Q}} \int_0^\infty \Bigg[ \frac{dV}{V_0} 
  \sum_{i_1=1}^c \dots\sum_{i_N=1}^{c} \frac{\prod_{i=1}^c{N_i!}}{N!}  \nonumber\\
&\quad \quad  \times e^{-\beta{\cal{F}}\left[T,V,\{N_i\}_{i=1,\dots,c}\right]-\beta PV+\beta\sum\limits_{j=1}^{N}(\tilde{\mu}_{i_j}-\tilde{\mu}_1)}B \Bigg],
\label{eq:Bmoy}
\end{align}
which we will evaluate using a MC sampler (Section~\ref{sec:mc_code}). All the microphysics of the system is contained in the partition function $\cal{Z}$. Our model for $\cal{Z}$ is detailed in Section~\ref{sec:theory}.

\subsection{A Simple Application of the Clapeyron Integration Method}
\label{sec:gd_validation}

As a simple example of the Clapeyron integration method, we now use an analytic plasma model for a BIM to evaluate the right-hand side of Eq.~\eqref{eq:clapeyron} and map the phase diagram of a two-component plasma. The purpose of this application is to illustrate the integration procedure and to show that the Clapeyron integration technique can reproduce exactly the same results as those obtained using more conventional techniques when the same input physics is assumed. An application of the full-fledged Clapeyron integration technique (using isobaric semi-grand canonical MC simulations) is presented in Section~\ref{sec:CO}.

We use the BIM model described by Ogata \textit{et al.} \citep{ogata1993} to compute the phase diagram of a C/O plasma. We choose this particular BIM model for this exercise as Ogata \textit{et al.} have published a C/O phase diagram based on this model to which our results can be compared. To allow a direct comparison with their results, we assume that the volume change during the phase transition is negligible, meaning that Eq.~\eqref{eq:clapeyron} simplifies to
\begin{equation}
\frac{d \beta}{d \xi_2} = \frac{\left( x_2^{\ell} - x_2^{s} \right)}{\xi_2 \left( 1 - \xi_2 \right) \left(u^{\ell} - u^s \right)},
\label{eq:clapeyron_vfix}
\end{equation}
where $u=U/N$ and where we have also fixed $\mu_1^{\rm ref}=\mu_2^{\rm ref}=0$.
To evaluate the right-hand side of Eq.~\eqref{eq:clapeyron_vfix} and integrate along the melting line, we need to extract $x_2$ and $u$ in the liquid and solid phases from the BIM model. The energy is obtained using a linear mixing rule of the one-component plasma (OCP) energies for which accurate fits to MC calculations already exist,
\begin{equation}
u = x_1 u^{\rm OCP} (\Gamma_1) + x_2 u^{\rm OCP} (\Gamma_2) + \Delta u^{\rm BIM} (R_Z, x_2, \Gamma_1),
\end{equation}
where $\Gamma_i = \frac{(Z_i e)^2}{a_i k_B T}$, $a_i = \left( \frac{3 Z_i}{4 \pi n_e} \right)^{1/3}$, and $R_Z = Z_2/Z_1$. The OCP terms are evaluated using Eq.~(11) of Ogata \textit{et al.} for the liquid (see also ref.~\cite{ogata1987}) and using their Eq.~(21) for the solid (see also ref.~\cite{dubin1990}). As for the correction term $\Delta u^{\rm BIM}$ to the linear mixing rule, we use the fits provided by Eqs.~(12) and (20) of Ogata \textit{et al.} 

The calculation of the concentrations requires a relation between $\xi_2$ and $x_2$. From the definition of the fugacity and fugacity fraction [Eqs.~\eqref{eq:fugacity_def} and \eqref{eq:fugacityfrac_def}], we find
\begin{equation}
\frac{\xi_2}{1-\xi_2} = e^{\beta ( \mu_2 - \mu_1)}.
\end{equation}
From $\mu_2 - \mu_1$, $x_2$ can then be obtained by numerically solving
\begin{equation}
\frac{\partial {\cal F}(N,N_2,V,T)}{\partial N_2} = \mu_2 - \mu_1.
\end{equation}
The Helmholtz free energy $\cal F$ is evaluated using the analytic fits provided by Ogata \textit{et al.} (their Eqs. 16, 25, 27, and 28),
\begin{equation}
{\cal F} = x_1 F^{\rm OCP} (\Gamma_1) + x_2 F^{\rm OCP} (\Gamma_2) + \Delta F^{\rm BIM}  (R_Z, x_2, \Gamma_1).
\end{equation}

With those equations in hand, we have everything we need to determine $x_2$ and $u$ in the liquid and solid phases and evaluate Eq.~\eqref{eq:clapeyron_vfix}. In what follows, we define species $i=1$ as C and $i=2$ as O. To start the integration of Eq.~\eqref{eq:clapeyron_vfix}, we must first specify an initial coexistence point $(\xi_{\rm O}^0,\beta^0)$. We choose to start the integration at $\xi_{\rm O}=0$, where only C ions are present in the plasma. The temperature $\beta^0$ at this coexistence point is then given by the melting temperature of the OCP, $\Gamma_m \simeq 175$ \citep{potekhin2000}. 

As $\frac{d \beta}{d \xi_2}$ is undefined in Eq.~\eqref{eq:clapeyron_vfix} for $\xi_2 = 0$, the first derivative has to be computed by other means. In applications of the Clapeyron integration technique to neutral systems (e.g., Lennard--Jones fluids), this initial derivative can be obtained through the infinite dilution limit and Henry's law \cite{mehta1994,hitchcock1999}. Here, this method demands the evaluation of $\langle \exp \left( - \beta \Delta u_{{\rm C} \rightarrow {\rm O}} \right) \rangle_{\rm NPT}$, where $\Delta u_{{\rm C} \rightarrow {\rm O}}$ represents the energy change that would result from the transformation of a C ion into an O ion. In practice, for dense plasmas, evaluating this term using MC simulations is challenging due to the strong fluctuations of the energy change $\Delta u_{{\rm C} \rightarrow {\rm O}}$. This issue is reminiscent of the limitations that affect particle insertion methods when applied to strongly coupled systems. Although this problem does not apply to the present section (as we are modeling the plasma using an analytic model), we still use the workaround that we have developed for our full MC-based Clapeyron integration (Section~\ref{sec:CO}). We initially assume that the temperature at the first $\xi_{\rm O}>0$ integration step ($\xi_{\rm O}^1$) is the same as that at $\xi_{\rm O}=\xi_{\rm O}^0=0$, i.e., $ \left. d \beta / d \xi_{\rm O} \right|_{\xi_{\rm O}^0}=0$. Using Eq.~\eqref{eq:clapeyron_vfix} and the BIM model, this allows us to get a first estimate of $ \left. d \beta / d \xi_{\rm O} \right|_{\xi_{\rm O}^1}$ and we then approximate the initial derivative as $ \left. d \beta / d \xi_{\rm O} \right|_{\xi_{\rm O}^0} =  \left. d \beta / d \xi_{\rm O} \right|_{\xi_{\rm O}^1}$. We then use this improved estimate of the initial derivative to obtain a refined estimate of the temperature at $\xi_{\rm O}^1$ and repeat this procedure until the derivative $\left. d \beta / d \xi_{\rm O} \right|_{\xi_{\rm O}^1}$ converges to a stable value.

Now that we have specified the initial coexistence condition and its initial derivative, the integration of Eq.~\eqref{eq:clapeyron_vfix} can begin. We define a grid of $\xi_{\rm O}$ values and use it to step from one $\xi_{\rm O}$ to the next. As the temperature at the next $\xi_{\rm O}$ value is initially unknown, we use a predictor--corrector algorithm to gradually refine its value until it stops varying by more than a fraction $\gamma$ of its value at the previous iteration. We refer the reader to ref.~\citep{hitchcock1999} for a detailed description of this algorithm. For each step, the $T$, $x_{\rm O}^{\ell}$, and $x_{\rm O}^s$ values are saved. After integrating all the way to $\xi_{\rm O}=1$, the phase diagram is directly given by the relation between those temperatures and concentrations. More specifically, $T(x_{\rm O}^{\ell})$ corresponds to the liquidus and $T(x_{\rm O}^{s})$ to the solidus. 

\begin{figure}
\includegraphics[width=\columnwidth]{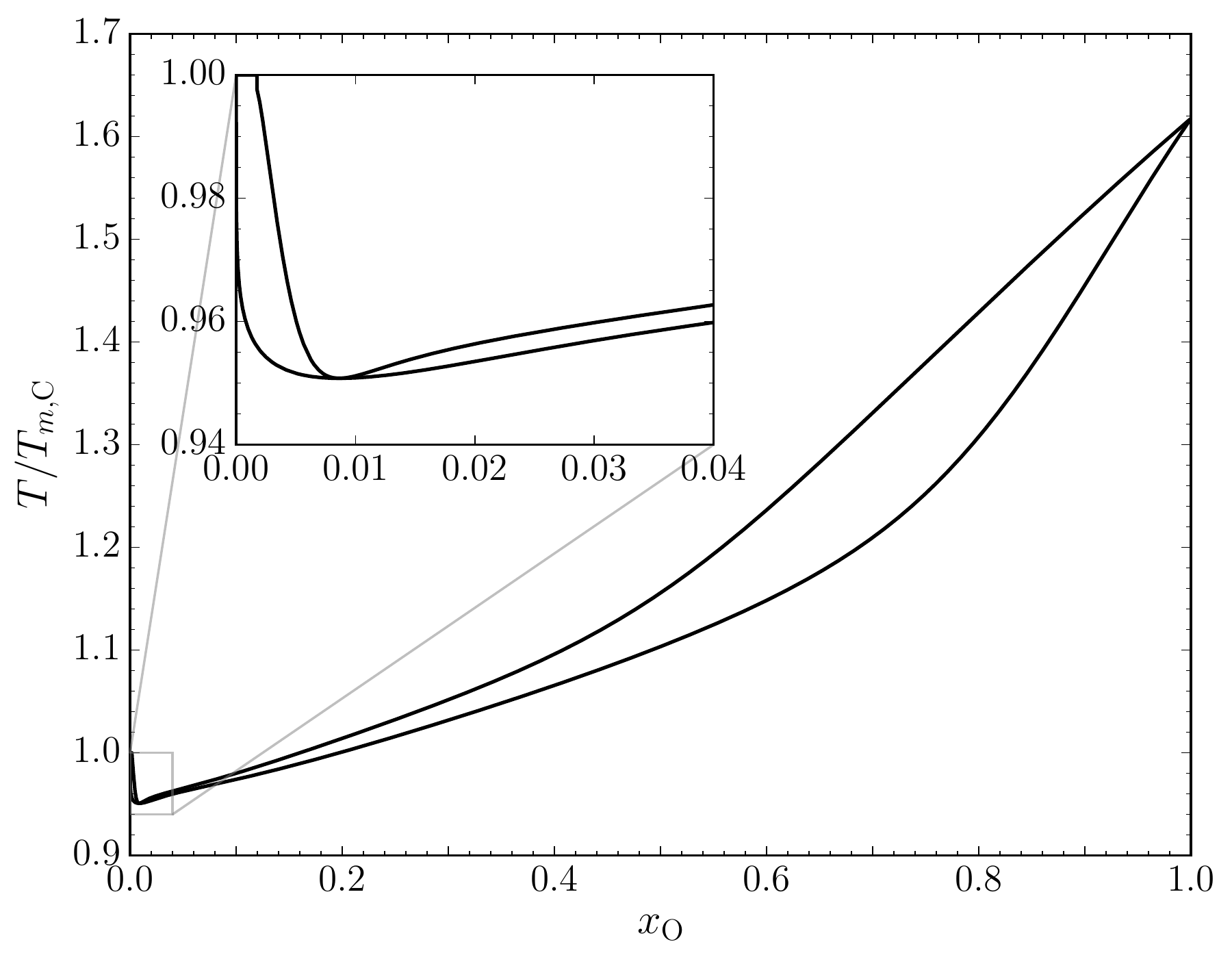}
\caption{C/O phase diagram computed using the Clapeyron integration technique and the BIM model of Ogata \textit{et al.} \citep{ogata1993} The horizontal axis is the O number concentration and the vertical axis is the ratio between the temperature and the melting temperature of a pure C OCP ($\Gamma=175$). The upper curve corresponds to the liquidus, and the lower one is the solidus.}
\label{fig:gb_ogata}
\end{figure}

Fig.~\ref{fig:gb_ogata} displays the resulting C/O phase diagram, which is almost identical to Fig.~5a of Ogata \textit{et al.} \citep{ogata1993}. The general shape as well as the position of the azeotropic point are the same. The only slight difference concerns the spurious behavior of our liquidus at very small O concentrations (see the inset of Fig.~\ref{fig:gb_ogata}). We attribute this difference to Ogata \textit{et al.}'s fit of $\Delta u_{\rm ex}^{\rm BIM}$ for the liquid, which is known to lead to unphysical results at small O concentrations \citep{dewitt1996,dewitt2003}.

\begin{figure}
\includegraphics[width=\columnwidth]{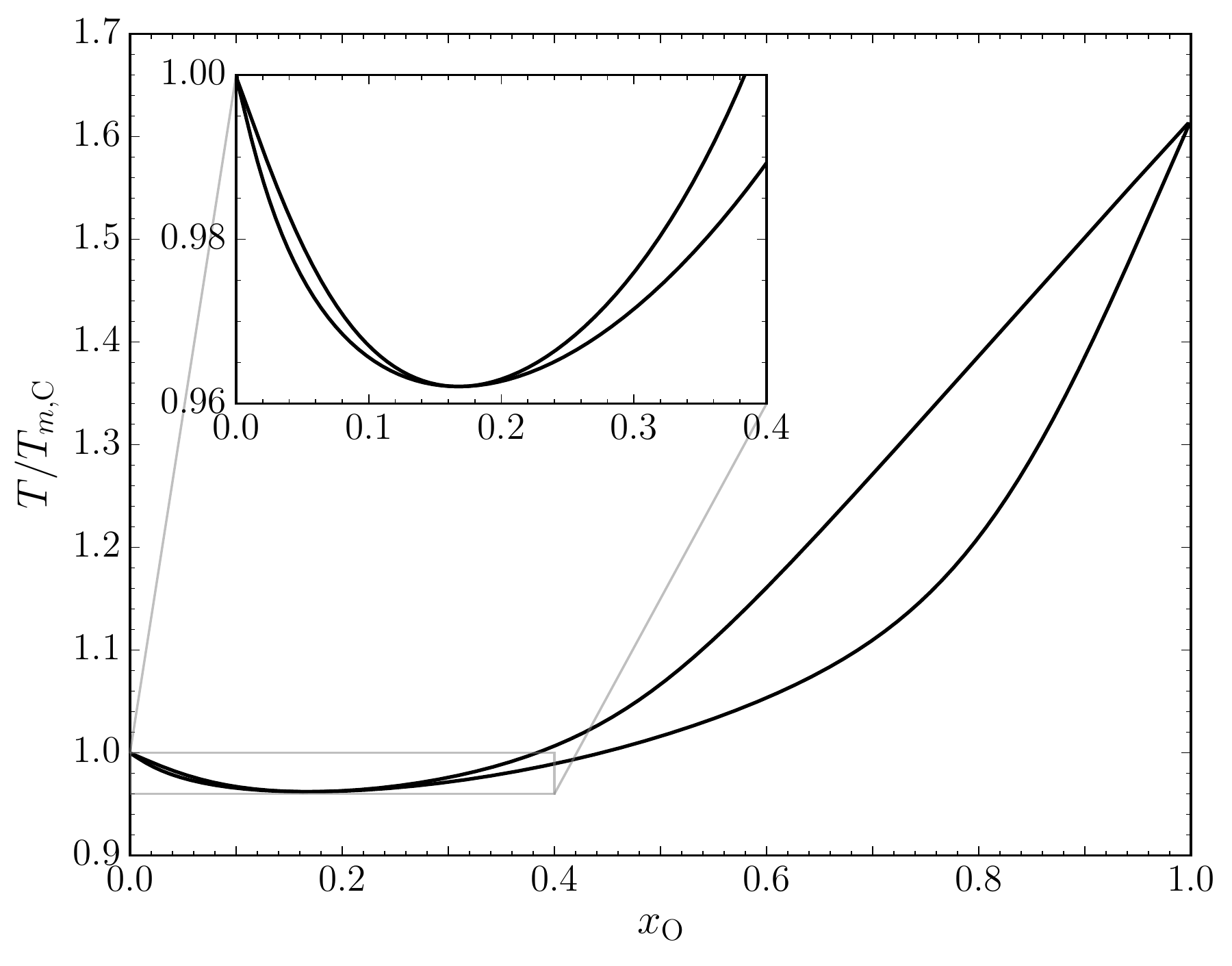}
\caption{Same as Fig.~\ref{fig:gb_ogata}, but assuming $\Delta u^{\rm BIM}=0$ for the liquid phase.}
\label{fig:gb_medin}
\end{figure}

We also tested what happens if we fix $\Delta u^{\rm BIM}=0$ for the liquid phase, which corresponds to the prescription adopted by Medin \& Cumming \citep{medin2010}. This simplification can be justified by the fact that $\Delta u^{\rm BIM}$ is very small in the liquid phase compared to the other energy terms. It also eliminates the spurious behavior of Ogata \textit{et al.}'s fit. With this approximation, we are able to reproduce the azeotropic phase diagram of Medin \& Cumming (compare Fig.~\ref{fig:gb_medin} to their Fig.~5a). We also replicated their finding that the phase diagram transitions from an azeotrope shape to a spindle shape when the charge ratio $R_Z$ goes below $\simeq 1.2$. This result is to be contrasted with the findings of ref.~\citep{dewitt1996}, who use different analytic fits and find that this transition occurs near $R_Z = 1.4$. This difference is very important in the context of white dwarf interiors ($R_Z = 1.33$ for a C/O plasma), where the shape of the phase diagram determines the composition profile of the frozen core. This comparison stresses the sensitivity of the final results on the fits used to derive the phase diagram. It clearly highlights the advantage of the Clapeyron integration method, which, once used in conjunction with isobaric semi-grand canonical MC simulations (Section~\ref{sec:CO}), requires no interpolation between simulation results.

\section{Plasma Model}
\label{sec:theory}
So far, our discussion has been general in the sense that no model for our plasma has been assumed. The microphysics is all contained in the canonical partition function $\cal{Z}$ [see Eqs.~\eqref{eq:calQ} and \eqref{eq:calF}]. We now specify a model for the electron--ion plasma appropriate for the conditions in white dwarf cores. Note, however, that the Clapeyron integration method is by no means limited to this particular model.

We consider a fully ionized plasma mixture of $c$ ionic species defined as in Section~\ref{Sec_II_B}. The system is contained in a volume $V=L^3$ with periodic boundary conditions in all three spatial directions. For notational simplicity, the charge (mass) of ion $J$ is denoted $Z_J$ ($M_J$), with $J=1,\dots,N$. We assume classical ions and quantum electrons. Approximating the ions as classical particles is well justified for the conditions in which we are interested. At the onset of crystallization in white dwarf cores, the interparticle distance is larger that the thermal de Broglie wavelength $\Lambda$. More specifically, $\Lambda/a \simeq 0.1 - 0.3$ where the liquid and solid phases coexist in white dwarf interiors.

\begin{figure}
\includegraphics[width=\columnwidth]{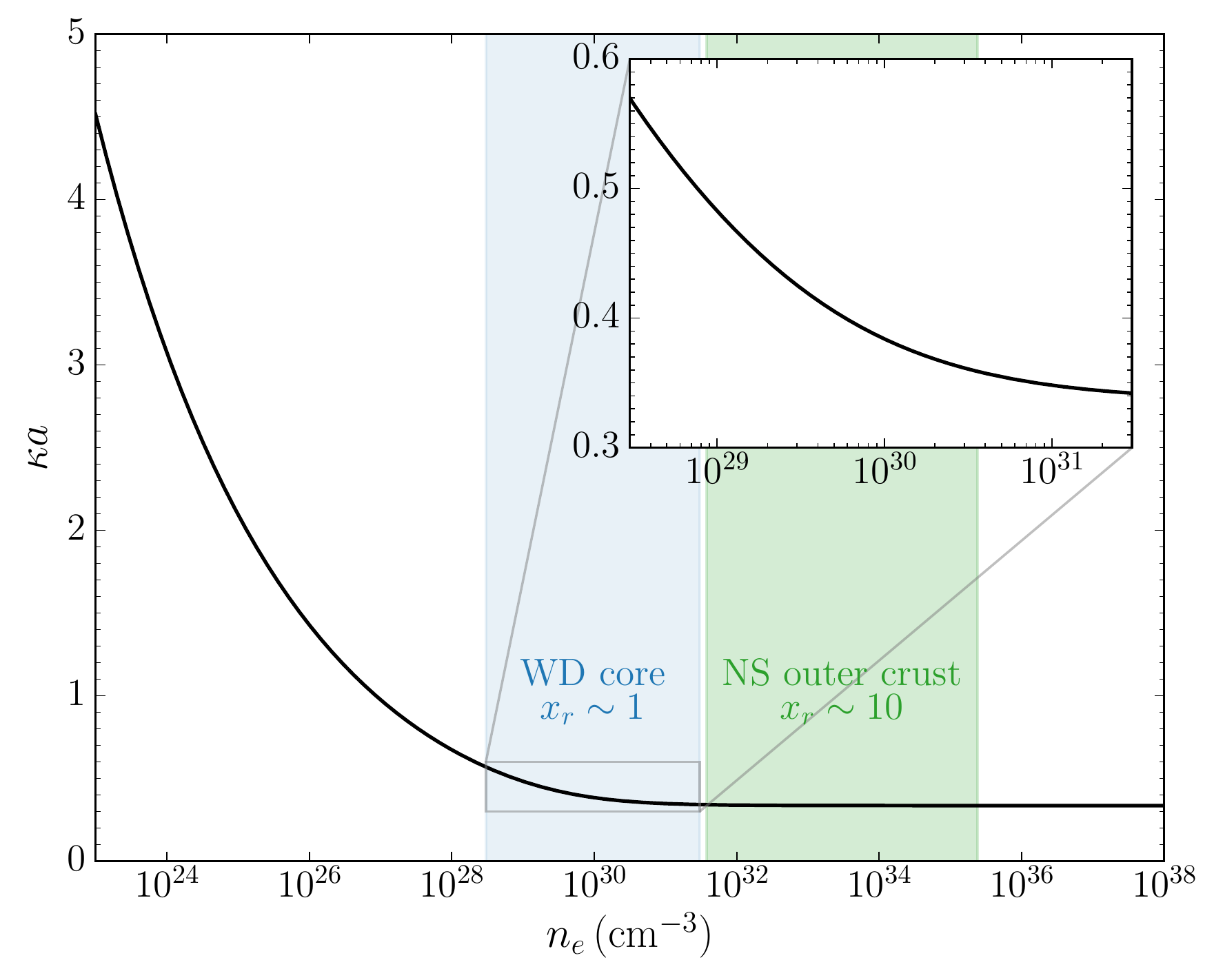}
\caption{Ratio of the average interparticle distance, $a=\left( 3Z/4\pi n \right)^{1/3}$, to the screening length, $1/\kappa$, as a function of the electronic density for a fully ionized C plasma [Eq.~\eqref{eq:kappa}]. The electrons are assumed to form a completely degenerate electron gas. $x_r=\frac{\hbar k_F}{m_e c}$ (with $k_F= (3 \pi^2 n_e)^{1/3}$ the Fermi momentum) is the relativistic parameter, and typical white dwarf (WD) core and neutron star (NS) crust conditions are highlighted. Note that the long-wavelength approximation (Appendix~\ref{sec:yukawa}) used here to evaluate $\kappa$ is no longer valid for the lower densities shown in this figure. It is nevertheless an excellent approximation for the dense astrophysical plasmas that are the focus of this work.}
\label{fig:kappa}
\end{figure}

Ion--ion, electron--ion, and electron--electron interactions need to be taken into account. We include electron--ion interactions to the lowest order, which yields (Appendix~\ref{sec:partfunc})
\begin{equation}
{\cal{Z}}\left(T,V,\{N_i\}_{i=1,\dots,c}\right)=\frac{1}{\prod_{i=1}^{c}{N_i! \Lambda_i^{3N_i}}}\int{dR^{3N} e^{-\beta{\cal{U}}(R^{3N})}}.
 \label{eq:calZ_model}
\end{equation}
Eq.~\eqref{eq:calZ_model} is the classical partition function of the ions interacting through the effective interaction energy
\begin{equation}
{\cal{U}}(R^{3N}) = {\cal{U}}_{\kappa}(R^{3N})+ F_{\rm jel}[n_e,T] \label{eq:calU}.
\end{equation}
The first term here is the potential of the system of ions interacting through the Yukawa (or screened Coulomb) interaction (which we justify in Appendix~\ref{sec:yukawa}), and the second term is the Helmholtz free energy of a relativistic homogeneous electron gas modeled at density $n_e=N_e/V$ and temperature $T$ (see Appendix~\ref{sec:partfunc}). For a pair of ions with charges $Z_i$ and $Z_j$, the Yukawa interaction potential takes the form
\begin{equation}
v_{\kappa} (r) = \frac{Z_i Z_j e^2}{4 \pi \epsilon_0} \frac{e^{-\kappa r}}{r},
\end{equation}
where $\epsilon_0$ is the vacuum permittivity and $1/\kappa$ is the relativistic Thomas--Fermi screening length. Fig.~\ref{fig:kappa} illustrates how this screening parameter varies as a function of the electronic density. Assuming this interaction potential, it follows that the potential of the system of interacting ions is given by
\begin{align}
{\cal{U}}_{\kappa}(R^{3N})&=\frac{e^2}{2\epsilon_0 V}\sum_{{\bf
    k},{\bf k}\neq 0}{{\vphantom{\sum}} \frac{1}{{\bf k}^2+\kappa^2}\left\{n_i({\bf
      k})n_i(-{\bf
      k})-\sum_{I=1}^N{Z_I^2}\right\}} \nonumber \\
&+\frac{e^2}{8\pi\epsilon_0}\left(\sum_{I=1}^N{Z_I^2}\right)
\left(E_\kappa-\kappa \right) \label{U_kappa},
\end{align}
where $n_i({\bf k})=\sum_{J=1}^{N}{Z_J e^{i{\bf k}\cdot{\bf R}_{J}}}$
is the Fourier transform of the ion charge density and $E_\kappa$ is the Madelung energy for the Yukawa interaction (see Appendix~\ref{sec:partfunc}). 

In contrast with this approach, most previous studies of plasma phase diagrams do not explicitly include the electronic background. This is a natural choice when working in the canonical (NVT) ensemble, but it is not appropriate when working in an isobaric ensemble. Indeed, when the volume and number of electrons are not fixed (as it is the case in the isobaric semi-grand canonical on which our Clapeyron integration approach is based), the background electronic energy is not a fixed quantity and it becomes necessary to include it explicitly. In any case, it is more rigorous to include the complete system of ions and electrons.

\section{Monte Carlo Implementation}
\label{sec:mc_code}
\subsection{General Overview of the MC Sampler}
We now describe how, given this plasma model, Eq.~\eqref{eq:Bmoy} can be evaluated using a MC sampler. Using Eqs.~\eqref{eq:fugacityfrac_def} and \eqref{eq:calZ_model}, the partition function [Eq.~\eqref{eq:calQ}] reads as
\begin{align}
&{\cal{Q}}\left(T,P,N, \{\xi_i\}_{i=2,\dots,c}\right)=\frac{1}{N!}\int_0^\infty \Bigg[ \frac{dV}{V_0} \sum_{i_1=1}^c \dots\sum_{i_N=1}^{c} \nonumber \\
&\quad \int \frac{dR^{3N}}{V^N} e^{-\beta{\cal{U}}(R^N) -\beta PV+N\ln \frac{V}{\Lambda_1^3}+\sum\limits_{j=1}^{N}\left(\ln\frac{\xi_{i_j}}{\xi_1}+\beta\mu_{i_j}^{\rm ref}\right)} \Bigg].
\label{eq:Qcal3}
\end{align}
To reach the targeted $P$ and $\{\xi_i\}_{i=2,\dots,c}$ conditions at a given $T$, three types of moves are required in isobaric semi-grand canonical MC simulations: (1) particles displacements, (2) volume changes, and (3) identity changes. From Eq.~\eqref{eq:Qcal3}, it follows that a particle displacement $R\to R'$, volume change $V\to V'$, or ionic identity change $i\to i'$ is accepted with probability 
\begin{align}
{\rm min}(1,e^\chi) \label{minoneexplambda},
\end{align}
where
\begin{align}
\chi=&-\beta ({\cal{U}}^\prime-{\cal{U}})-\beta P(V^\prime-V)+N\ln\frac{V'}{V} \nonumber\\
&+\ln\frac{\xi_{i^\prime}}{\xi_i}+\beta (\mu_{i^\prime}^{\rm ref}-\mu_{i}^{\rm ref}).
\label{eq:lambda}
\end{align}
The role of the reference chemical potential $\mu_i^{\rm ref}$ that we have introduced earlier [Eq.~\eqref{eq:muref_first}] now becomes apparent. If we assume that $\mu_i^{\rm ref}=0$ for all $i=1,...,c$, then $\chi$ is dominated by the change in the electron free energy, i.e., $\chi \simeq -\beta(F[n_e^\prime,T]-F[n_e,T])$. We have $n_e^\prime-n_e=(Z_{i^\prime}-Z_i)\frac{N}{V}$ and typically $\chi \simeq -\beta (Z_{i^\prime}-Z_i)\mu_e$ with $\mu_e$ the electron chemical potential. Under degenerate conditions, $\mu_e \propto n_e^{2/3}$ and therefore $\left| \chi \right| \gg 1$ in dense plasmas (under white dwarf conditions, $n_e \sim 10^{30}\,{\rm cm}^{-3}$). Then, the acceptance probability ${\rm min}(1,e^\chi)$ is either 1 or 0 and eventually all but one ionic species disappear from the MC simulation. In other words, the mapping $x_i \leftrightarrow \xi_i$ is not practical: all the variation in $x_i$ occurs over a small range of $\xi_i$ values very close to 0 or 1. The reference chemical potentials were introduced to overcome this difficulty. We define them as
\begin{equation}
\mu_i^{\rm ref} = Z_i \mu_e^{\rm ref},
\end{equation}
where $\mu_e^{\rm ref}$ is fixed to a value close to $\mu_e [n_e,T]$. With this choice, the dominant $\beta (Z_{i^\prime}-Z_i)\mu_e$ term in Eq.~\eqref{eq:lambda} is largely cancelled by the $\beta (\mu_{i^\prime}^{\rm ref}-\mu_{i}^{\rm ref})$ term and the $x_i \leftrightarrow \xi_i$ mapping becomes more practical (i.e., $x_i \sim \xi_i$). Note that the electronic density needed to define $\mu_e^{\rm ref}$ is a priori unknown, as the electronic density fluctuates during the MC simulation. An iterative process is therefore needed in order to find the value of $\mu_e^{\rm ref}$ that leads to a well-behaved $x_i \leftrightarrow \xi_i$ mapping (once selected, $\mu_e^{\rm ref}$ is kept fixed during the integration of the phase diagram). 

At each iteration during the MC simulation, the algorithm decides randomly whether a particle displacement, volume change, or identity change is attempted (and Eq.~\ref{minoneexplambda} is then evaluated to decide whether the move is actually accepted). A standard prescription is to assign probabilities of $N/(2N+1)$, $1/(2N+1)$ and $N/(2N+1)$ to attempting a particle displacement, volume change, and identity change, respectively (with this prescription, $2N+1$ attempts represent one MC cycle) \cite{hitchcock1999}. Empirically, we found that assigning a probability of $90\%$ to ion displacements, $5\%$ to volume changes, and $5\%$ to identity changes leads to a much quicker convergence of the MC simulation. While this choice affects the particular trajectory of the MC simulations, we have verified that it does not influence the average energies, concentrations, and densities extracted from the simulations.

Typically, a few thousand MC cycles are needed before the targeted pressure and fugacity fractions are reached. After that, a few more thousand cycles are performed in order to accurately evaluate the average concentrations and enthalpies needed for the integration of the Clapeyron equation (uncertainties are estimated using the block-averaging technique \citep{flyvbjerg1989,grossfield2009}). A series of tests designed to validate our MC code are presented in Appendix~\ref{sec:validation}.

\subsection{Numerical Implementation of the MC Sampler}
\label{sec:mc_details}
Consider a given ionic configuration ${\bf R}=R^{3N}$ of the $N$ ions in a cubic simulation box of volume $V=L^3$ where periodic conditions are imposed on all boundaries. The MC algorithm necessitates the calculation of the energy ${\cal{U}}_{\kappa}({\bf R})$ (Eq.~\ref{U_kappa}), and of the all the interparticle forces $-\frac{\partial{\cal{U}}_{\kappa}({\bf R})}{\partial {\bf R}}$, which are used to calculate the instantaneous contribution to the total pressure. More specifically, we have
\begin{equation}
P=-\frac{\partial {\cal{F}}}{\partial V}=P_i+P_e,
\end{equation}
where the electronic pressure is given by $P_e=-\frac{\partial F_{\rm jel}}{\partial V}$ and, using Eqs.~\eqref{eq:calF} and \eqref{eq:calZ_model}, the ionic pressure is
\begin{equation}
P_i=nk_BT-\frac{1}{3V}\left\langle {\bf R}\cdot \frac{\partial{\cal{U}}_{\kappa}}{\partial {\bf R}}\right\rangle -\frac{1}{3V}\left\langle L\frac{\partial {\cal{U}}_{\kappa}}{\partial L}\right\rangle.
\end{equation}

Because the bare Coulomb interactions between ions are only weakly shielded by the electrons ($\kappa a\sim 0.35$, see Fig.~\ref{fig:kappa}), the range of the Yukawa potential is large in the sense that one cannot safely truncate the potential at the distance $r=L/2$ and make use of the usual minimum image convention and neglect the interaction of a particle with the particles in the periodically replicated cells.
To overcome this problem, we use the Ewald summation technique to evaluate the sums over ${\bf n}\in\mathbb{Z}^3$ in Eq.~\eqref{U_kappa} (e.g., see ref.~\cite{frenkel2002}).
For a Yukawa potential (e.g., ref.~\cite{salin2000}), the interaction energy $v_\kappa(r)=\frac{e^2}{4\pi\epsilon_0}\frac{e^{-\kappa r}}{r}$ between two particles at distance $r$ is represented by the sum of a short-range (sr) and a long-range (lr) component,
\begin{equation}
v(r)=\phi_{\rm sr}(r)+\phi_{\rm lr}(r),
\end{equation}
with
\begin{align}
\phi_{\rm sr}(r)&=\frac{e^2}{8\pi\epsilon_0 r}\left[\erfc\left(\alpha r+\frac{\kappa}{2\alpha}\right)e^{\kappa r}\right. \nonumber \\
&\quad \left.+\erfc\left(\alpha r-\frac{\kappa}{2\alpha}\right)e^{-\kappa r}\right]
\end{align}
and
\begin{equation}
\phi_{\rm lr}(r)\!=\!\!\frac{e^2}{\epsilon_0 V}\sum_{{\bf n}\in\mathbb{Z}^3}{\frac{e^{-(k^2+\kappa_{sc}^2)/(4\alpha^2)}}{k^2+\kappa_{sc}^2} e^{i{\bf k}\cdot{\bf r}}},
\end{equation}
where ${\bf k}=\frac{2\pi}{L}{\bf n}$, $\alpha>0$ is the Ewald parameter (a numerical parameter conveniently chosen to optimize the evaluation of the previous expressions \cite{frenkel2002}), and $\erfc$ is the complementary error function. In our simulation code, the Ewald sum is numerically evaluated with the particle--particle--particle--mesh ($\rm P^3M$) method, which combines high resolution of close encounters (the sr term is calculated using nearest-neighbor techniques) and rapid long-range force calculations (the lr forces are computed on a mesh using three-dimensional fast Fourier transforms) \cite{frenkel2002}. The code is fully parallelized using Message Passage Interface (MPI). Compared to the standard implementation of the $\rm P^3M$ algorithm, here, MC simulations in the isobaric semi-grand canonical ensemble require carefully reinitializing the algorithm to account for the changes in the simulation box size and screening length $1/\kappa$ that occur whenever a volume change or a particle identity change is performed.

\section{Application: the C/O interior of white dwarfs}
\label{sec:CO}

\subsection{C/O Phase Diagram}
We now combine the MC code presented in the previous Section with the Clapeyron integration algorithm described in Section~\ref{sec:gd} to obtain the phase diagram of a dense C/O plasma under white dwarf conditions. As in Section~\ref{sec:gd_validation}, we start our integration from a pure C plasma and we define species $i=1$ as C and $i=2$ as O. We perform the integration of Eq.~\eqref{eq:clapeyron} on a grid defined by $\xi_{\rm O} = \{0,0.02,0.04,\dots,1\}$ and we fix the pressure to $10^{18}$~bar, a typical pressure for white dwarf cores. Other numerical parameters are listed in Table~\ref{tab:num_params}.

\begin{table}
\caption{\label{tab:num_params} Numerial parameters used for our integration of the C/O phase diagram.}
\begin{ruledtabular}
\begin{tabular}{lr}
Parameter & Value\\
\hline
Total number of ions in each MC simulation & 686\\
Number of MC cycles per simulation & 6000\\
$\mu_{e}^{\rm ref}$ & 523.7\,keV\\
Predictor--corrector convergence criterion ($\gamma$) & 0.001 \\
\end{tabular}
\end{ruledtabular}
\end{table}

\begin{figure}
\includegraphics[width=\columnwidth]{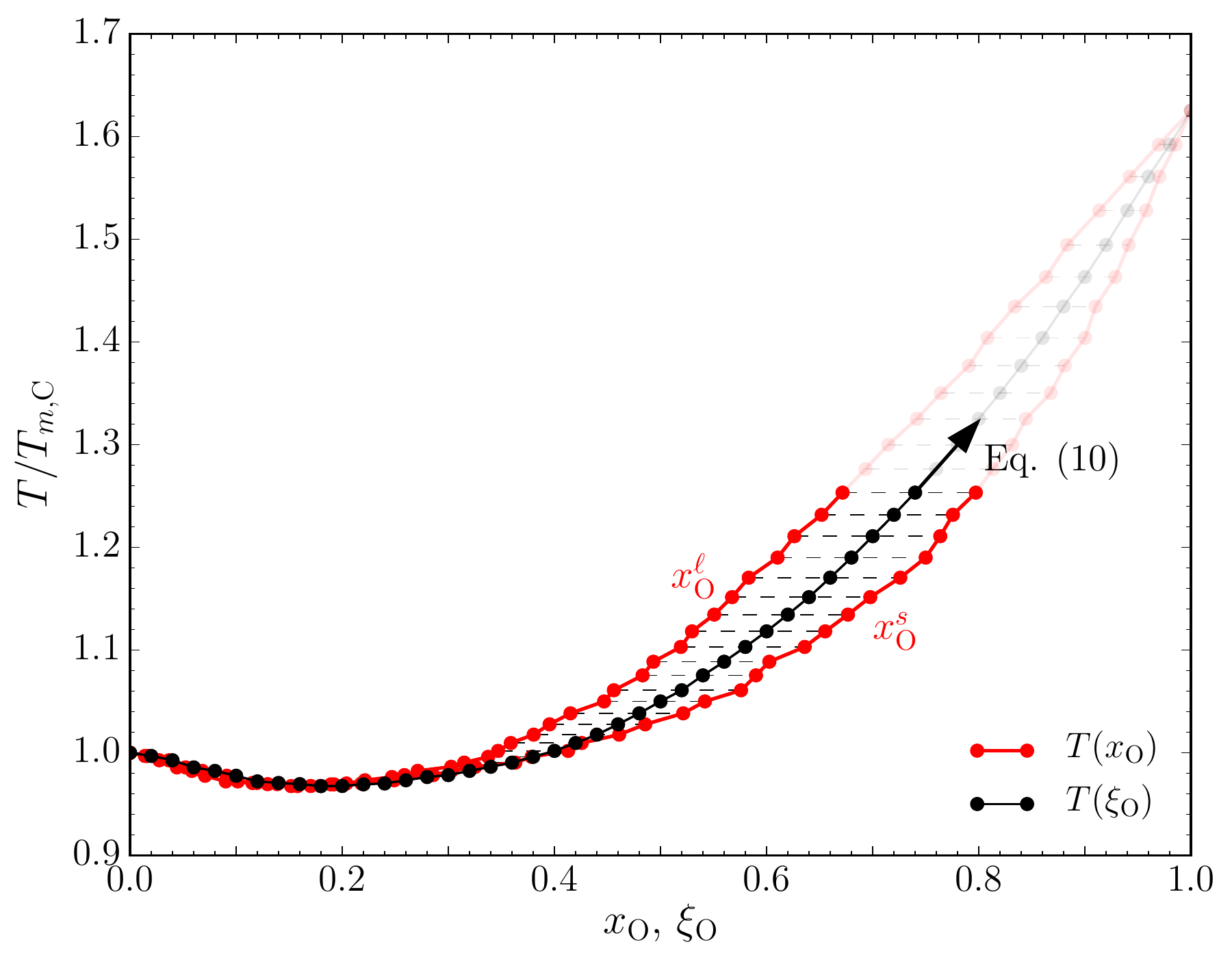}
\caption{Clapeyron integration of the C/O phase diagram. The black line shows the $T(\xi_{\rm O})$ coexistence line that we integrate from $\xi_{\rm O}=0$ to 1 using Eq.~\eqref{eq:clapeyron}. Each point along the $T(\xi_{\rm O})$ coexistence line yields an O concentration in the liquid ($x_{\rm O}^{\ell}$) and in the solid ($x_{\rm O}^{s}$) phases, tracing the liquidus and the solidus, respectively. The temperature is shown in units of the melting temperature of a pure C Yukawa plasma (here, $\Gamma=178$). Note that $T(\xi_{\rm O})$ depends on our choice of $\mu_e^{\rm ref}$ ($523.7\,{\rm keV}$ here), but the $T(x_{\rm O})$ lines do not.}
\label{fig:CO_explain}
\end{figure}

\begin{figure}
\includegraphics[width=\columnwidth]{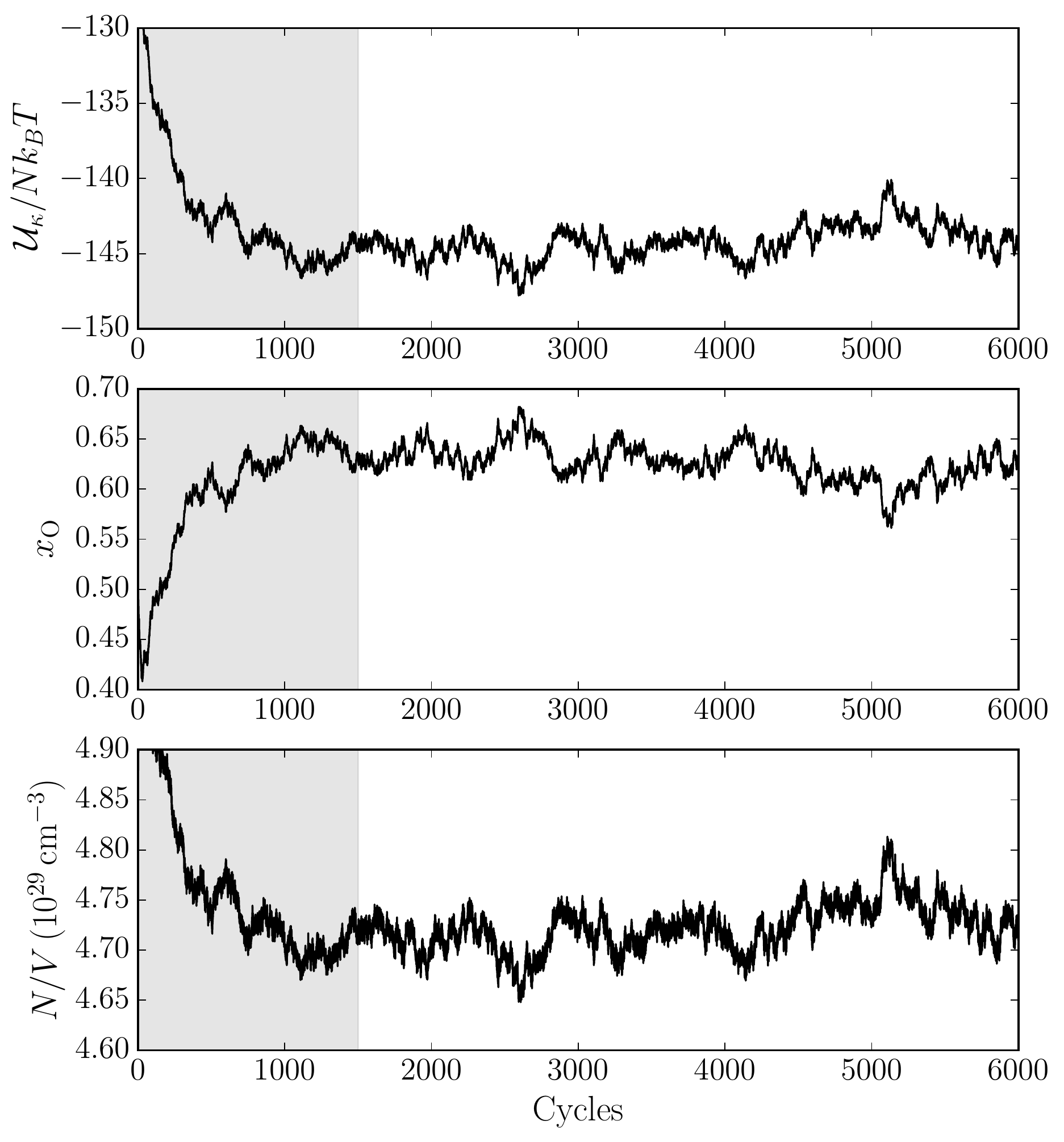}
\caption{Evolution of the ion excess energy (${\cal U}_{\kappa}/N k_B T$), O number fraction ($x_{\rm O}$), and ion density ($N/V$) during one the MC simulations used to map the C/O phase diagram. For this simulation, the plasma is in the liquid phase, $\xi_{\rm O}=0.7$, $T=470.48\,{\rm eV}$, and $P=10^{18}$\,bar. Other numerical parameters are specified in Table~\ref{tab:num_params}. The region in gray corresponds to the equilibration phase that we ignore when we compute the average enthalpies and concentrations needed to evaluate the Clapeyron equation.}
\label{fig:CO_MC_demo}
\end{figure}

For the initial coexistence condition $\left( \xi_{\rm O}^0, \beta^0 \right)$, we use the melting temperature given in ref.~\cite{hamaguchi1996} for Yukawa systems (here, $\kappa a \simeq 0.35$, which implies $\Gamma_m \simeq 178$). The initial derivative $\frac{d \beta}{d \xi_{\rm O}}$ is computed as in Section~\ref{sec:gd_validation} and the numerical integration of Eq.~\eqref{eq:clapeyron} is performed using the predictor--corrector algorithm described in ref.~\cite{hitchcock1999}. For each $\xi_{\rm O}$ value, the liquid and solid phases are simulated simultaneously (and independently), yielding a pair of concentrations $(x_{\rm O}^{\ell},x_{\rm O}^{s})$ for each $T(\xi_{\rm O})$ point along the coexistence line (see Fig.~\ref{fig:CO_explain}). The evolution of one of those MC simulations is shown in Fig.~\ref{fig:CO_MC_demo}. For the solid phase, the C and O ions are randomly positioned on a bcc lattice and their displacements are limited in order to prevent the solid from melting. The bcc phase is the only solid phase accessible to Yukawa systems near the OCP limit (large screening length $1/\kappa$) such as those found in white dwarf interiors \cite{hamaguchi1996}. 

\begin{figure}
\includegraphics[width=\columnwidth]{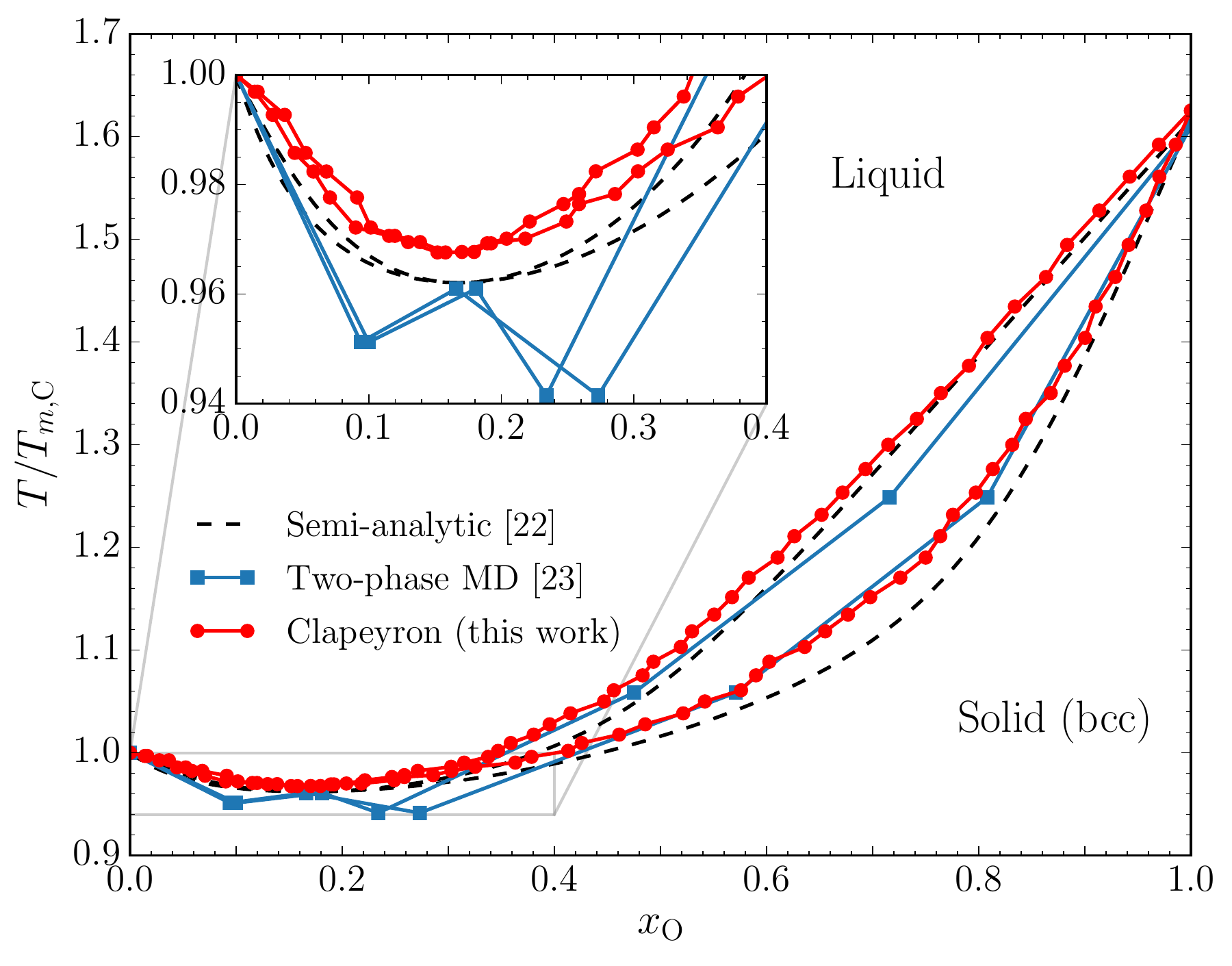}
\caption{C/O phase diagram obtained via integration of the Clapeyron relation [Eq.~\eqref{eq:clapeyron}] using our plasma model described in Section~\ref{sec:theory} (in red). For comparison, we also show the C/O phase diagrams of refs.~\cite{medin2010,horowitz2010}. The temperature is shown in units of the melting temperature of a pure C Yukawa plasma (here, $\Gamma=178$). A similar figure was presented in ref.~\cite{blouin2020}.}
\label{fig:CO}
\end{figure}

Fig.~\ref{fig:CO} shows the resulting C/O phase diagram. The smoothness of our phase diagram illustrates the high level of accuracy achieved by our MC simulations. Moreover, we recover the melting temperature of a pure O plasma, $T_{m,{\rm O}}$, at the end of our integration at $\xi_{\rm O}=x_{\rm O}=1$,  $T_{m,{\rm O}} = \left[Z({\rm O}) / Z({\rm C}) \right]^{5/3} T_{m,{\rm C}} \simeq 1.62 T_{m,{\rm C}}$. This result is not explicitly enforced by the Clapeyron integration method. It can only be achieved if the integration from $\xi_{\rm O}=0$ to $\xi_{\rm O}=1$ is accurate enough to recover this known limit.

As pointed out in the companion paper \cite{blouin2020}, our C/O phase diagram is close to that of Medin \& Cumming \cite{medin2010} (dashed lines in Fig.~\ref{fig:CO}). Both have a similar azeotrope shape, with azeotropic points at about the same concentrations. This is a remarkable result given the approximations underlying their phase diagram. Namely, the ion--ion interactions are not screened and the excess energy of the liquid phase is assumed to simply be the sum of the OCP energies of each ionic component (but note that Medin \& Cumming explore the impact of deviations from this linear mixing rule in their Appendix~B). Our C/O phase diagram is also similar to that of Horowitz \textit{et al.} \cite{horowitz2010}. Apart from the superior sampling made possible by our relatively inexpensive method, the main difference is that the separation between the liquidus and the solidus, $\Delta x_{\rm O}$, is slightly larger in our case (at concentrations higher than the azeotrope). As briefly discussed by Horowitz \textit{et al.}, $\Delta x_{\rm O}$ could be underestimated in their simulations due to finite-size effects that cause an artificial composition gradient across the liquid--solid interface. We can expect that their phase diagram would converge to something closer to ours if they used larger MD simulations.

The results shown in Fig.~\ref{fig:CO} were obtained at $P=10^{18}\,{\rm bar}$. We also computed additional versions of this phase diagram assuming different pressures. Consistent with our findings for the OCP (Appendix~\ref{sec:validation}), we found that the resulting phase diagram remains practically unchanged for the range of pressures that characterize white dwarf interiors.

\subsection{Analytic Fits to our C/O Phase Diagram}
\label{sec:fits}
\begin{figure}
\includegraphics[width=\columnwidth]{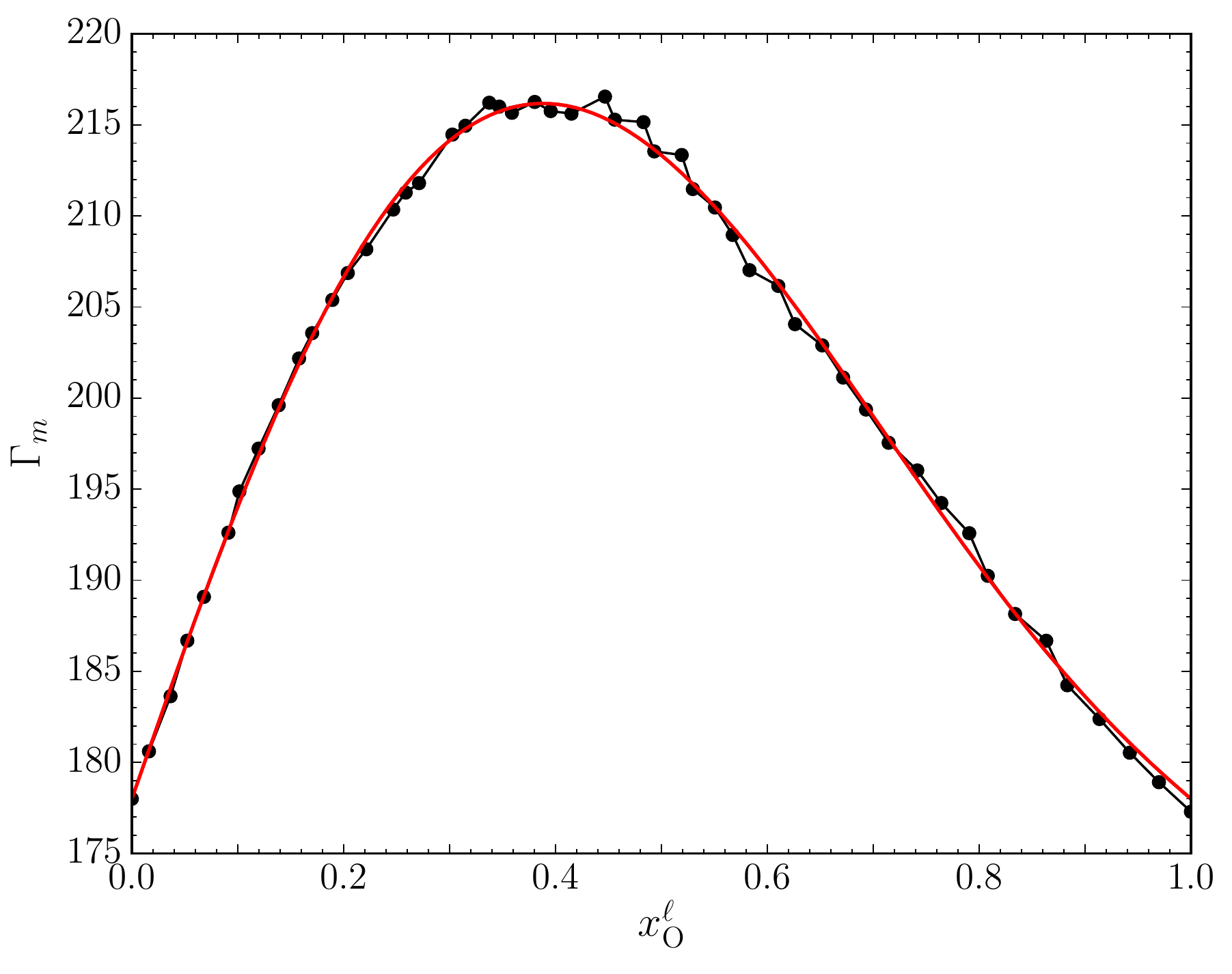}
\caption{Coulomb coupling parameter at which the C/O liquid with a concentration $x_{\rm O}^{\ell}$ crystallizes. Results from our Clapeyron integration are shown in black and the analytic fit given by Eq.~\eqref{eq:fit} is in red.}
\label{fig:gammam_fit}
\end{figure}

\begin{figure}
\includegraphics[width=\columnwidth]{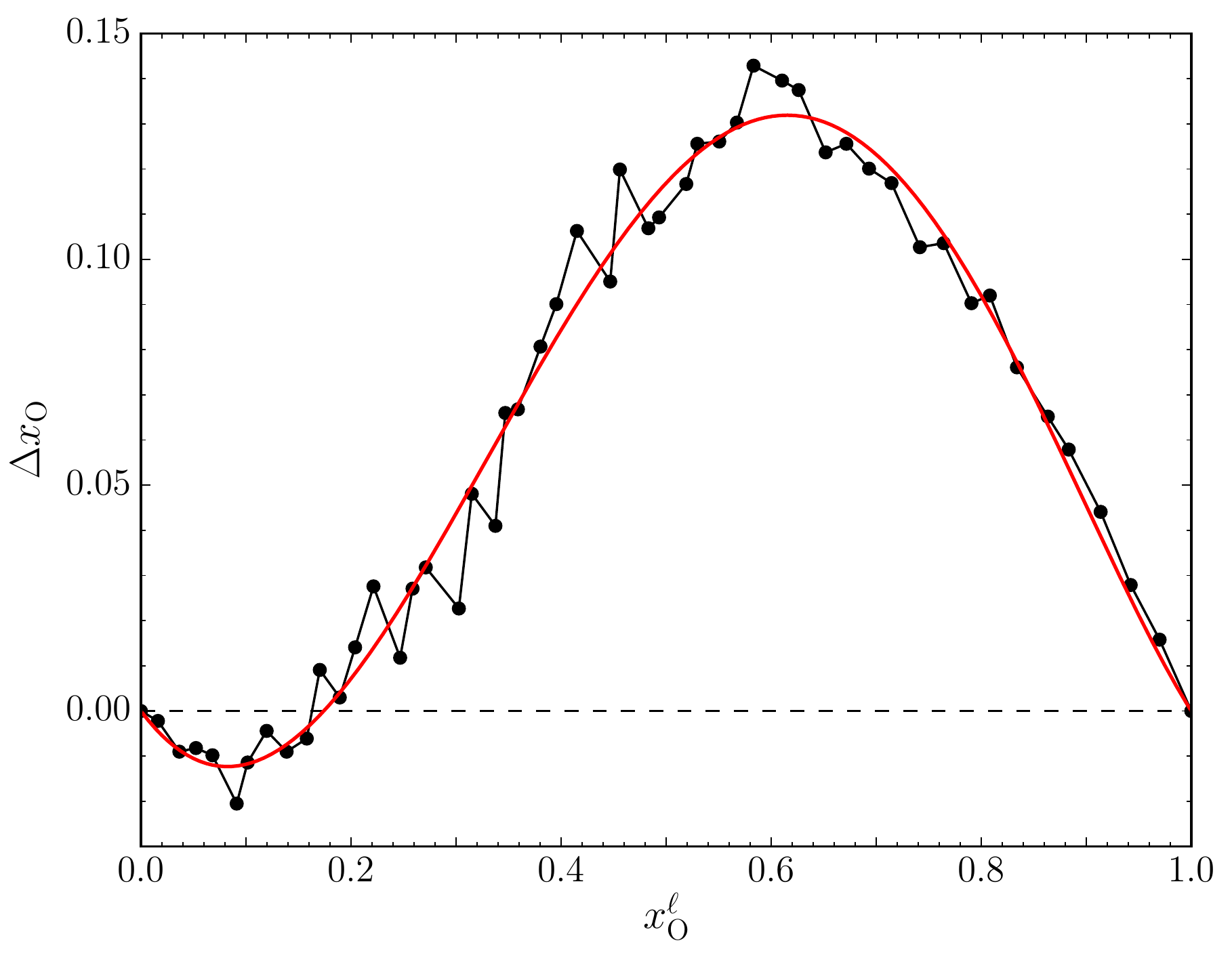}
\caption{Difference between the O concentration of the coexisting solid and liquid phases as a function of the O concentration of the liquid at the phase transition. In black, we show the results taken directly from Fig.~\ref{fig:CO} and, in red, we show our analytic fit [Eq.~\eqref{eq:fit}].}
\label{fig:deltax_fit}
\end{figure}

To facilitate the implementation of our phase diagram in white dwarf codes, we provide analytic fits to the Coulomb coupling parameter at the melting temperature, $\Gamma_m (x_{\rm O}^{\ell})$, and to the separation between the liquidus and the solidus, $\Delta x_{\rm O} (x_{\rm O}^{\ell})$. The coupling parameter of the mixture is computed as
\begin{equation}
\Gamma=\frac{\langle Z ^{5/3} \rangle e^2}{a_e k_B T},
\end{equation}
with $\langle Z^{\alpha} \rangle = \sum_i Z_i^{\alpha} n_i /n$ and $a_e = (3/ 4\pi n_e)^{1/3}$.
Both $\Gamma_m (x_{\rm O}^{\ell})$ and $\Delta x_{\rm O} (x_{\rm O}^{\ell})$ can be accurately fitted with a fifth-order polynomial,
\begin{equation}
\sum_{i=0}^5 a_i (x_{\rm O}^{\ell})^i,
\label{eq:fit}
\end{equation}
where the coefficients $a_i$ are given in Table~\ref{tab:fit}. Our fit to $\Gamma_m (x_{\rm O}^{\ell})$ is shown in Fig.~\ref{fig:gammam_fit}. Note that the fit recovers the known limits $\Gamma_m (x_{\rm O}^{\ell}=0) = \Gamma_m (x_{\rm O}^{\ell}=1) = 178$ for a one-component Yukawa system with a screening parameter typical of white dwarf interiors ($\kappa a \simeq 0.35$) \cite{horowitz2010,hamaguchi1996,vaulina2002}. Fig.~\ref{fig:deltax_fit} shows our fit to $\Delta x_{\rm O} (x_{\rm O}^{\ell})$. The fit is such that $\Delta x_{\rm O} (x_{\rm O}^{\ell})=0$ in the pure C and pure O limits, as well as at the azeotropic point ($x_{\rm O}^{\ell} \approx 0.18$). Both fits reproduce the simulations accurately within the statistical noise.

\begin{table}
\caption{\label{tab:fit} Fit parameters for $\Gamma_m (x_{\rm O}^{\ell})$ and $\Delta x_{\rm O} (x_{\rm O}^{\ell})$ [Eq.~\eqref{eq:fit}].}
\begin{ruledtabular}
\begin{tabular}{lrr}
 & $\Gamma_m (x_{\rm O}^{\ell})$ & $\Delta x_{\rm O} (x_{\rm O}^{\ell})$\\
\hline
$a_0$ & 178.000000 & 0.000000 \\ 
$a_1$ & 167.178104 & $-$0.311540\\ 
$a_2$ & $-$3.973461 & 2.114743\\ 
$a_3$ & $-$741.863826 & $-$1.661095\\ 
$a_4$ & 876.516929 & $-$1.406005 \\ 
$a_5$ & $-$297.857813 & 1.263897 \\ 
\end{tabular}
\end{ruledtabular}
\end{table}

\section{Conclusion}
\label{sec:conclusion}
We have presented how the Clapeyron integration method, in conjunction with isobaric semi-grand canonical MC simulations, can be used to map the phase diagrams of dense multicomponent plasmas. This technique has many advantages compared to competing approaches: (1) all calculations are performed directly at the coexistence conditions (no analytic fits required and no uninteresting state points to simulate); (2) no phase transition interface needs to be modeled (thereby greatly simplifying the mitigation of finite-size effects); (3) no particle insertions/deletions are required; (4) all thermodynamic properties of the system at the phase transition are available at no additional cost; (5) the underlying MC simulations allow a fine sampling of the coexistence line at a reasonable cost; (6) the electronic background is explicitly included; and (7) all calculations are performed at constant pressure, as is appropriate for phase equilibrium. As an example application, we have computed the phase diagram of a dense C/O plasma under conditions relevant for white dwarf interiors. Our results are in good agreement with previous calculations and we have provided analytic fits to facilitate the implementation of this new, accurate C/O phase diagram in existing white dwarf evolution codes. 

This paper has focused on applications to dense, two-component electron--ion plasmas. However, the Clapeyron integration method can in principle be used for a much wider range of systems. In planetary science for instance, it could prove useful to tackle problems such as the demixing of H/He mixtures in the interiors of giant planets \citep{morales2009,lorenzen2009} or the melting temperature of iron in Earth's core \citep{laio2000,alfe2009}. In the near future, we plan to apply this technique to $c>2$ component dense plasmas and to generalize it to metallic alloys.

\acknowledgements
The authors are indebted to Didier Saumon for insightful discussions from which both the conception and execution of this project have benefited. Research presented in this article was supported
by the Laboratory Directed Research and Development program of Los
Alamos National Laboratory under project number 20190624PRD2.
This work was performed under the auspices of the U.S. Department of Energy
under Contract No. 89233218CNA000001.

\appendix

\section{Clapeyron Equation for a Three-Component System}
\label{sec:3component}
Let us start from the Gibbs--Duhem relation under its fugacity form [Eq.~\eqref{eq:gd2}]. Assuming that $P$, $T$, and $N$ are fixed and remembering that the liquid--solid coexistence conditions imply that $f_i^{\ell}=f_i^s$ and $\xi_i^{\ell}=\xi_i^s$ for all $i=1,2,3$, we have the relation
\begin{equation}
\sum_{i=1}^{3}{x_i^{\ell}\frac{d\xi_i}{\xi_i}}=\sum_{i=1}^{3}{x_i^{s}\frac{d\xi_i}{\xi_i}}
\end{equation}
for all coexistence states. Since by definition $x_1=1-x_2-x_3$ and $\xi_1=1-\xi_2-\xi_3$, we find the Clapeyron equation
\begin{equation}
\left.\frac{d\xi_2}{d\xi_3}\right|_{P,T,N}=-\frac{\xi_2(1-\xi_2)(x_3^{\ell}-x_3^{s})+\xi_2\xi_3(x_2^{\ell}-x_2^{s})}{\xi_3(1-\xi_3)(x_2^{\ell}-x_2^{s})+\xi_2\xi_3(x_3^{\ell}-x_3^{s})}.
\label{eq:clapeyron3c}
\end{equation}
To begin the integration of Eq.~\eqref{eq:clapeyron3c}, prior knowledge of one coexistence point $\xi_2(\xi_3)$ for a given $T$ and $P$ is needed. A practical way to do this is to first map the two-component phase diagram of species $i=1,2$ for a given $P$ [using Eq.~\eqref{eq:clapeyron}], which would directly yield $\xi_2(\xi_3=0)$ for a range of $T$ along the coexistence line for a fixed $P$. Then, the integration of Eq.~\eqref{eq:clapeyron3c} can start from this initial $\xi_2(\xi_3=0)$ coexistence point and proceed along a grid of $\xi_3$ values from 0 to 1. Since $x_1=1-x_2-x_3$, each state point along the coexistence line defined by Eq.~\eqref{eq:clapeyron3c} yields a pair of compositions $(x_1^{\ell},x_2^{\ell},x_3^{\ell})$ and $(x_1^{s},x_2^{s},x_3^{s})$. The integration of Eq.~\eqref{eq:clapeyron3c} can then be repeated for different $T$ in order to map the three-component phase diagram in the full temperature--composition space. Note that the integrations at different $T$ are independent from one another and can be performed simultaneously. We have successfully applied this approach to the three-component C/O/Ne mixture found in white dwarf cores \cite{blouin2021}.

\section{Model Partition Function}
\label{sec:partfunc}
To derive our partition function [Eq.~\eqref{eq:calZ_model}], we start from the total Hamiltonian of the ions and electrons. This is essential to keep track of all the terms that are important when working in the isobaric semi-grand canonical ensemble. The Hamiltonian of the ions and electrons in a cubic volume $V=L^3$ with periodic boundary conditions is conveniently split in the form (see, e.g., p.\,409 of ref.~\cite{hansen1986} for $c=1$, here generalized to $c>1$)
\begin{equation}
H=H_{i}+H_e+H_{ei}.
\end{equation}
$H_{i}$ is the Hamiltonian of all ions immersed in a homogeneous neutralizing background,
\begin{align}
H_{i}&=\sum_{I=1}^N\frac{{\bf P}_I^2}{2M_I} \nonumber\\
&\quad +\frac{1}{2V}\sum_{{\bf n}\in\mathbb{Z}^3}{{\vphantom{\sum}}' v({\bf k})\left[n_i({\bf k})n_i(-{\bf k})-\sum_{I=1}^N{Z_I^2}\right]}\\
&\quad +\frac{e^2}{8\pi\epsilon_0}\left(\sum_{I=1}^N{Z_I^2}\right) E \nonumber
\end{align}
where ${\bf k}=2 \pi {\bf n} /L$ with ${\bf n}\in\mathbb{Z}^3$, the prime means ${\bf n}\ne 0$, $v({\bf k})=\frac{e^2}{\epsilon_0}\frac{1}{k^2}$ is the Fourier transform of the Coulomb potential, $n_i({\bf k})=\sum_{J=1}^{N}{Z_J e^{i{\bf k}\cdot{\bf R}_{J}}}$ is the Fourier transform of the ion charge density, and $E$ is the Madelung energy,
\begin{equation}
\frac{e^2}{4\pi\epsilon_0} E=\lim_{r\to 0}\left(\frac{1}{V}\sum_{{\bf n}\in\mathbb{Z}^3}{{\vphantom{\sum}}' v({\bf k})e^{i{\bf k}\cdot{\bf r}}}-\frac{e^2}{4\pi\epsilon_0}\frac{1}{r}\right).
\end{equation}
$H_e$ is the Hamiltonian of all electrons immersed in a homogeneous neutralizing background (jellium model). Finally, $H_{ei}$ is the electron--ion interaction term,
\begin{equation}
H_{ei}=\frac{1}{V}\sum_{{\bf n}\in\mathbb{Z}^3}{{\vphantom{\sum}}' v({\bf k})\delta n_i({\bf k})\delta n_e(-{\bf k})},
\end{equation}
where $n_e({\bf k})=Z_e\sum_{j=1}^{N_e}{e^{i{\bf k}\cdot{\bf r}_j}}$ (with $Z_e=-1$) is the electron charge density and $\delta n_{i,e}({\bf r})=n_{i,e}({\bf r})-n_{e}$ (with $n_e=N_e/V$) is the fluctuating density.

For a system of classical ions and quantum electrons, the total canonical partition function is given by
\begin{multline}
{\cal{Z}}\left(T,V,\{N_i\}_{i=1,\dots,c}\right)=\\
\frac{1}{\prod_{i=1}^{c}{N_i! \,h^{3N_i}}}\int{dR^{3N}dP^{3N}{\rm
    Tr}_e e^{-\beta H}}, \label{calZ}
\end{multline}
where $R^{3N}$ and $P^{3N}$ represent the positions and momenta of ions, and ${\rm Tr}_e$ is the trace over a complete set of electronic states in the field due to a fixed ionic configuration $R^N$.
If $H_{ei}$ is treated as a perturbation (i.e., to lowest order in the interaction Hamiltonian $H_{ei}$), it is well established that the partition function can then be expressed as \cite{ashcroft1978,hansen1986},
\begin{multline}
{\cal{Z}}\left(T,V,\{N_i\}_{i=1,\dots,c}\right)=\\\frac{1}{\prod_{i=1}^{c}{N_i! \Lambda_i^{3N_i}}}\int{dR^{3N} e^{-\beta{\cal{U}}(R^{3N};n_e,T)}}.
\end{multline}
In other words, at this order, ${\cal{Z}}$ reads as the classical partition function of the system of ions interacting according to the effective interaction energy
\begin{multline}
{\cal{U}}(R^{3N};n_e,T))=\frac{1}{2V}\sum_{{\bf n}}{{\vphantom{\sum}}' v({\bf k})\left[\frac{n_i({\bf k})n_i(-{\bf k})}{\epsilon({\bf k})}-\sum_{I=1}^N{Z_I^2}\right]}\\
+\frac{e^2}{8\pi\epsilon_0}\left(\sum_{I=1}^N{Z_I^2}\right) E+F_{\rm
  jel}[n_e,T]. \label{eq:calU_1}
\end{multline}
Here, $F_{\rm jel}[n_e,T]=-k_BT\ln {\rm Tr}_ee^{-\beta H_e}$ is the free energy of the homogeneous, relativistic electron gas at density $n_e$ and temperature $T$. In this work, we evaluate this term as $F_{\rm jel} = F_{\rm id} + F_{\rm xc}$, where $F_{\rm id}$ is the free energy of the relativistic quantum ideal gas and $F_{\rm xc}$ is the exchange--correlation free energy. For $F_{\rm id}$, we use its well-known expression in terms of generalized Fermi--Dirac integrals, which we compute using the parametrization of ref.~\cite{gong2001}, and for $F_{\rm xc}$, we follow ref.~\cite{stolzmann1996}. In Eq.~\eqref{eq:calU_1}, $\epsilon({\bf k})$ is the dielectric function of the relativistic homogeneous electron gas model. As explained in Appendix~\ref{sec:yukawa}, for the physical conditions of interest in this work, the long-wavelength approximation $\epsilon(k)=1+\frac{\kappa^2}{k^2}$ (where $\kappa=k_{\rm TF} \left( 1 + x_r^2 \right)^{1/4}$ is the relativistic inverse screening length) is well justified. Under this approximation, Eq.~\eqref{eq:calU_1} is conveniently written as in Eqs.~\eqref{eq:calU} and \eqref{U_kappa} with the Madelung energy 
\begin{equation}
E_\kappa=\lim_{r\to 0}\left(\frac{1}{V}\sum_{{\bf
        n}}{{\vphantom{\sum}}' \frac{4\pi}{k^2+\kappa^2} e^{i{\bf k}\cdot {\bf r}}}-\frac{e^{-\kappa r}}{r}\right).
\end{equation}

\section{Yukawa Potential}
\label{sec:yukawa}

In the random-phase approximation, the static dielectric function of a relativistic electron gas
in its ground state is given by
\begin{equation}
\epsilon(k) = 1 + \frac{k_{\rm TF}^2}{k^2} Y(k,x_r),
\label{eq:epsilon_full}
\end{equation}
where $k_{\rm TF}$ is the Thomas--Fermi inverse screening length,
\begin{equation}
k_{\rm TF} = \frac{e \left( 12 \pi m_e n_e \right)^{1/2}}{\hbar k_F},
\end{equation}
$Y(k,x_r)$ is given in ref.~\cite{jancovici1962} (see also Appendix~B of ref.~\cite{daligault2009} for a more convenient form), 
$k_F= (3 \pi^2 n_e)^{1/3}$ is the Fermi momentum, and $x_r = \frac{\hbar k_F}{m_e c}$ is the relativistic parameter. For the very dense plasmas in which we are interested ($n_e \sim 10^{30}\,{\rm cm}^{-3}$ in white dwarf cores), the range of $k$ values for which $Y(k,x_r)$ is significantly different from $Y(k=0,x_r)$ occurs only at large $k$, when $k_{\rm TF}^2/k^2 \sim 0$. Therefore, the static dielectric function is well approximated by
\begin{equation}
\epsilon (k) = 1 + \frac{k_{\rm TF}^2}{k^2} Y(k=0,x_r) = 1 + \frac{\kappa^2}{k^2}
\end{equation}
and the screened interaction potential can be accurately described by a Yukawa potential. If we use the full expression for $\epsilon(k)$ [Eq.~\eqref{eq:epsilon_full}], we find a difference of $<0.3\%$ for white dwarf conditions between the resulting potential and the much simpler Yukawa potential, which justifies the long-wavelength approximation used above. Taking the limit $k \rightarrow 0$ of $Y(k=0,x_r)$, we obtain
\begin{equation}
\kappa = k_{\rm TF} \left( 1 + x_r^2 \right)^{1/4}.
\label{eq:kappa}
\end{equation}
The behavior of $\kappa$ as a function of the electron density is shown in Fig.~\ref{fig:kappa}. Note that under white dwarf conditions, we are not yet in the ultrarelativistic limit ($x_r \gg 1$) where $\kappa a$ becomes independent of the electron density.

\section{Validation of the MC Code}
\label{sec:validation}
In this Appendix, we present the different tests that we have performed to validate our MC code. We first benchmark our simulations in the canonical (NVT) ensemble. This is the simplest case, as only particle displacements are allowed in the MC simulations. Then, we add volume changes to our simulations and perform simulations in the NPT ensemble. After validating our NPT implementation, we use it to quantify the effect of volume changes during the liquid--solid phase transition of the OCP and of Yukawa systems. Finally, we present calculations in the isobaric semi-grand canonical ensemble (NPT$\Delta\mu$). This is the most complex case, where displacements, volume changes, and identity changes are allowed.

\subsubsection{NVT Ensemble}
We first verified that our code can reproduce the energy values reported in refs.~\cite{johnson1993} and \cite{farouki1994} for Lennard--Jones and Yukawa systems, respectively. Our energy values are all consistent (within the statistical uncertainties) with those reported in these works. 

\begin{figure}
\includegraphics[width=\columnwidth]{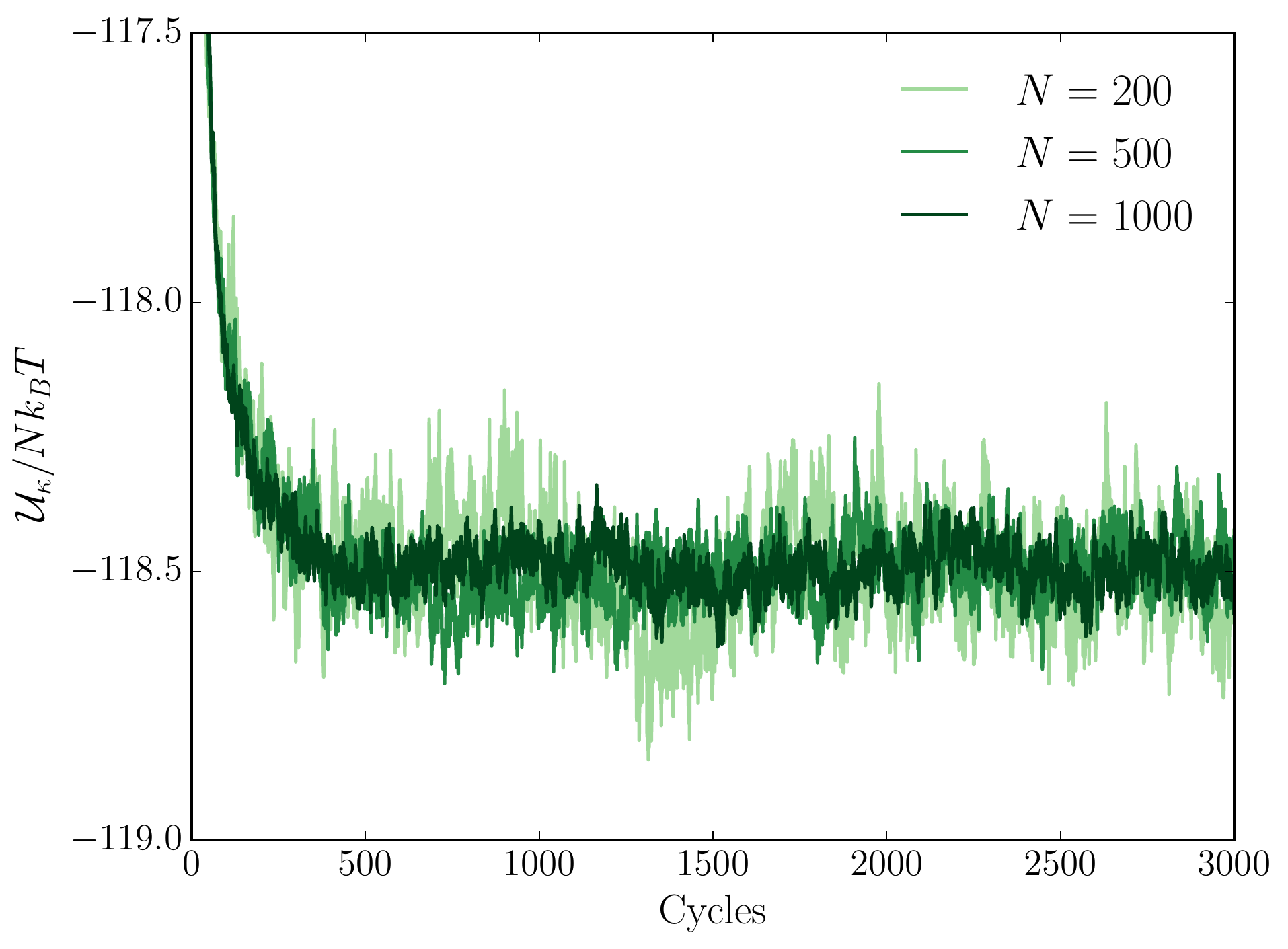}
\caption{Evolution of the excess ionic energy of a pure C plasma in the liquid phase for different system sizes. The ion density is $5 \times 10^{29}\,{\rm cm}^{-3}$ (which corresponds to $\kappa a =0.36$, see Fig.~\ref{fig:kappa}), $T=400\,{\rm eV}$, and NVT simulations are employed. We find a mean ${\cal{U}}_{\kappa}/N k_B T$ of $-118.517\pm0.021$, $-118.494 \pm 0.008$ and $-118.497\pm 0.006$ for the $N=200$, $N=500$ and $N=1000$ simulations, respectively. The first 1000 cycles (the equilibration phase) were excluded from the calculation of the averages. As elsewhere in this Appendix, the uncertainties correspond to $1\sigma$ confidence intervals.}
\label{fig:nvt_demo}
\end{figure}

In Fig.~\ref{fig:nvt_demo}, we show how the ion excess energy fluctuates during MC simulations of a pure C plasma (modeled as a Yukawa system) in the liquid phase using different system sizes. The physical conditions used here ($N/V= 5 \times 10^{29}\,{\rm cm}^{-3}$ and $T=400\,{\rm eV}$) are typical of those found in white dwarf interiors. Even for a very small system containing only 200 ions the energy converges to the same average value than larger systems, indicating that finite-size effects can be mitigated at minimal cost. However, smaller system sizes are associated with increased statistical fluctuations that decrease the precision of the mean. To consider this effect, we evaluate the uncertainties of MC simulation averages using the block-averaging technique throughout this paper.

As another validation of our NVT simulations, we evaluate the melting temperature, $\Gamma_m$, of the OCP ($\kappa=0$). To do so, we follow the approach of Hamaguchi \textit{et al.} (1996) \citep{hamaguchi1996}. More precisely, the Helmholtz free energy of each phase (liquid and bcc solid) is computed via thermodynamic integration on the Coulomb coupling parameter $\Gamma$ with
\begin{equation}
\frac{F^{\ell}(\Gamma)}{N k_B T} = \int_0^{\Gamma}{\frac{U_{\rm ions}(\Gamma)}{N k_B T} \frac{ d \Gamma'}{\Gamma'}  } + \frac{F_{\rm id} (\Gamma)}{N k_B T}
\end{equation}
and
\begin{equation}
\frac{F^{s}(\Gamma)}{N k_B T} = \int_{\infty}^{\Gamma}{\left[ \frac{U_{\rm th}(\Gamma)}{N k_B T} - \frac{3}{2} \right] \frac{ d \Gamma'}{\Gamma'}  } + \frac{F_{\rm harm} (\Gamma)}{N k_B T},
\end{equation}
where the ideal terms $F_{\rm id}$ and $F_{\rm harm}$ are given in Hamaguchi \textit{et al.}, $U_{\rm ions}$ is the ion excess energy (Eq.~\ref{U_kappa}), $U_{\rm th}$ is the thermal component of the ion excess energy (that is, the ion energy minus the Madelung energy), $\Gamma=Z^2 e^2 /(a k_B T)$, and $a=(3/4 \pi n)^{1/3}$. To perform the thermodynamic integrations, $U_{\rm ions}$ and $U_{\rm th}$ are modeled with the analytic interpolation functions (Eqs.~11 and 18 of Hamaguchi \textit{et al.} 1996, see also refs.~\cite{stringfellow1990,dubin1990})
\begin{equation}
\frac{U_{\rm ions} (\Gamma)}{N k_B T} = a \Gamma + b \Gamma^{1/3} + c + d \Gamma^{-1/3}
\label{eq:fit_uliq}
\end{equation}
and
\begin{equation}
\frac{U_{\rm th} (\Gamma)}{N k_B T} = \frac{3}{2} + \frac{A_1}{\Gamma} + \frac{A_2}{\Gamma^2},
\end{equation}
where $a$, $b$, $c$, $d$, $A_1$, and $A_2$ are free parameters that are adjusted to the energies extracted from our MC simulations. The NVT simulations used to compute those energies were performed with 1024 ions and for 4000 MC cycles. 

Once the $U_{\rm ions}$ and $U_{\rm th}$ fits are constructed, the melting temperature is obtained by finding the temperature $\Gamma_m$ where $F^{\ell}(\Gamma) = F^{s}(\Gamma)$. We find a melting temperature of $\Gamma_m = 174.6 \pm 1.6$, in excellent agreement with the modern $\Gamma_m = 175$ value \citep{potekhin2000}. The uncertainty on our $\Gamma_m$ value was found by propagating the statistical uncertainties of the MC energy values to the analytic fits, which results in a range of $\Gamma$ values where the liquid and solid free energies intersect.

\subsubsection{NPT Ensemble}
\begin{figure}
\includegraphics[width=\columnwidth]{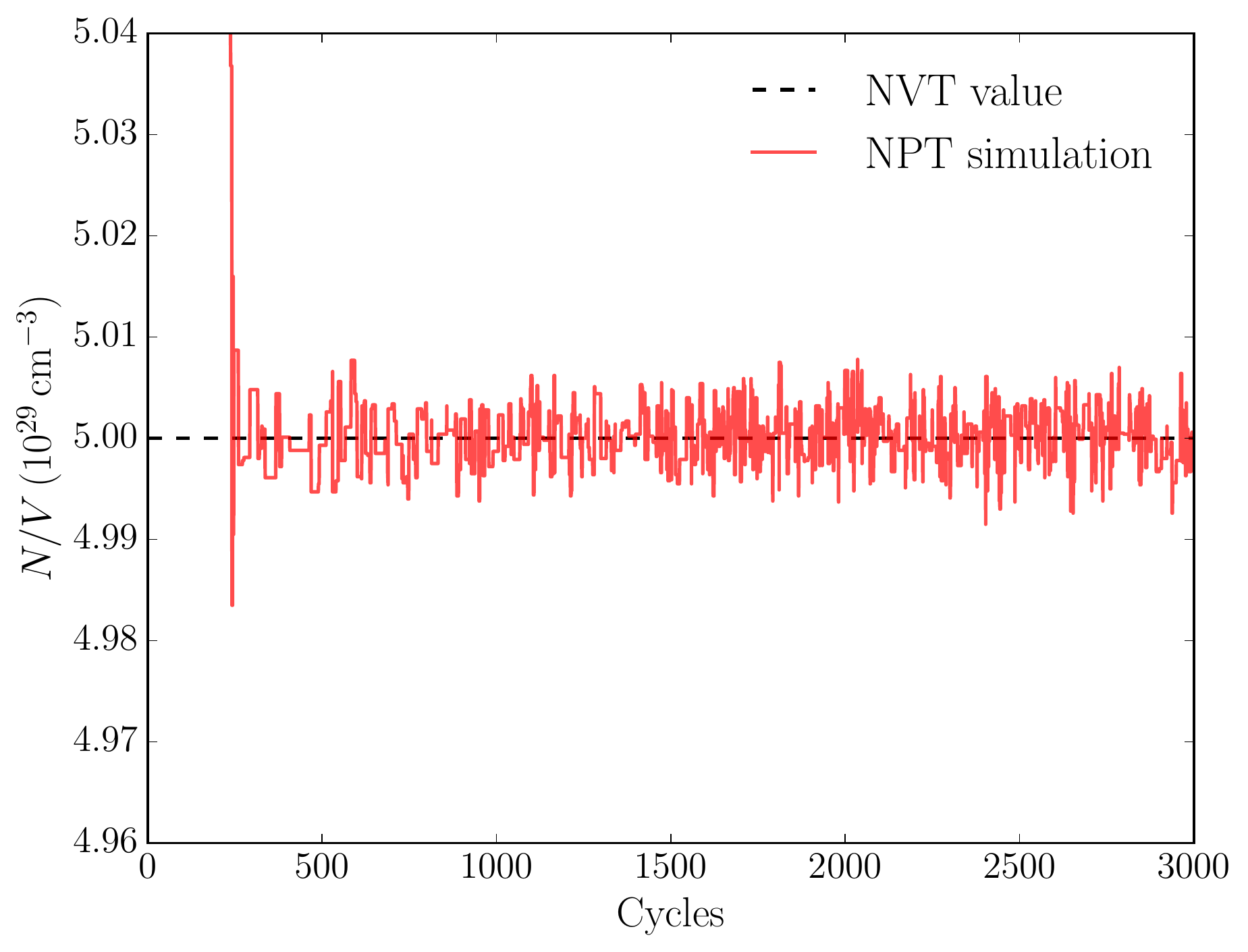}
\caption{Evolution of the ion density for a pure C plasma in the liquid phase simulated using an NPT simulation. The target pressure is $8.27 \times 10^{18}\,{\rm bar}$ (which corresponds to the pressure found using NVT simulations with an ion density of $5 \times 10^{29}\,{\rm cm}^{-3}$), $T=400\,{\rm eV}$, and the screening length $1/\kappa$ is dynamically adjusted according to the electron density. The average density in the NPT simulation is $(5.0003\pm0.0002) \times 10^{29}\,{\rm cm}^{-3}$, demonstrating that our NVT and NPT simulations are consistent.}
\label{fig:npt_demo}
\end{figure}

Now that we have thoroughly validated our NVT simulations, we add one layer of complexity and investigate NPT simulations. Testing NPT simulations is straightforward: for a given system and temperature, an NPT simulation should yield the same density (or volume) as that imposed to an NVT simulation if the pressure imposed to the NPT simulation corresponds to the pressure obtained in the NVT simulation. Fig.~\ref{fig:npt_demo} shows how this convergence is successfully achieved for NPT calculations performed under the same conditions as those used in Fig.~\ref{fig:nvt_demo}. Note that the convergence of the density can be significantly accelerated if the simulation box is initialized with a volume close to its final value.

Using our NPT simulations, we reevaluated the melting temperature of the OCP at constant pressure. Nearly all investigations of the liquid--solid phase transition of the OCP and of Yukawa systems assume that the volume change $\Delta V = V^{s} - V^{\ell}$ during the transition is negligible \citep{slattery1980,nagara1987,dubin1990,stringfellow1990,farouki1993,farouki1994,hamaguchi1996,hamaguchi1997}. In dense plasmas, this is usually justified by invoking the fact that the pressure of the degenerate electron gas completely dominates the total pressure \cite{hansen1977}. Using this approximation, the melting point---which normally corresponds to the temperature where the Gibbs free energies of both phases intersect---is simply given by the temperature where the Helmholtz free energies are equal, $F^{\ell} = F^{s}$. This assumption has at least two important practical advantages: (1) simulations can be performed at constant volume and (2) the electron jellium can be ignored since the electronic density is the same in both phases at the transition. Here, we investigate the validity of this approximation with a phase transition calculation that explicitly considers volume changes.

To compute the Gibbs free energy $G$ for given $P$ and $T$ conditions, we use a thermodynamic integration method where we vary the coupling parameter $\lambda$, which is simply a prefactor applied to the ion--ion interaction potential. For the liquid, the integration is performed between a reference state corresponding to the ideal gas ($\lambda = \Gamma = 0$) and the target state at $\lambda=1$,
\begin{equation}
g(\lambda=1) - g(\lambda=0) = \int_{0}^{1}{\frac{d \lambda' }{\lambda'} \langle u_{\rm ions}(\lambda') \rangle_{{\rm NPT}, \lambda'}},
\label{eq:npt_integ_liq}
\end{equation}
where $g = G / \left( N k_B T \right)$ and $u_{\rm ions} = U_{\rm ions}  / \left( N k_B T \right)$ is extracted from our MC simulations in the NPT ensemble for a given $\lambda$ value. The reference ideal term is given by
\begin{align}
G(N,P,T,\lambda=0) &= F_{\rm id,ions} (N,V,T)  \nonumber
\\&\quad + F_{\rm jel} [n_e,T] + PV,
\end{align}
where $F_{\rm id,ion} (N,V,T)$ is the ideal-gas free energy (see, e.g., Eq.~5 of Hamaguchi \textit{et al.} 1997 \citep{hamaguchi1997}), $F_{\rm jel} [n_e,T]$ is the electron jellium free energy, and the volume $V$ is found by numerically solving
\begin{equation}
\frac{N}{V} k_B T - \left. \frac{\partial F_{\rm jel}}{\partial V} \right|_{N,T} = P
\end{equation}
for a given total pressure $P$. Numerically, we integrate Eq.~\eqref{eq:npt_integ_liq} in two parts. For $\lambda\leq 0.1$, we use Simpson's rule to evaluate the integral using the results from 20 NPT simulations with $0<\lambda \leq 0.1$. Note that for $\lambda=0$ the integrand of Eq.~\eqref{eq:npt_integ_liq} is simply 0: in the limit of small coupling, $u_{\rm ions}(\lambda) \propto \lambda^{3/2}$ (see Eq.~18 of ref.~\cite{hansen1973}). For the the $0.1 < \lambda \leq 1$ interval, the simulations are fitted to the functional form of Eq.~\eqref{eq:fit_uliq} and the integration is performed analytically.

For the solid phase, the reference state is the harmonic system and we have
\begin{equation}
g(\lambda=1) - g_h = \int_{\infty}^{1} { \frac{d \lambda'}{\lambda'} \left\langle u_{\rm th} (\lambda') - \frac{3}{2} \right\rangle_{\rm{NPT}, \lambda'}}.
\label{eq:npt_int_solid}
\end{equation}
The harmonic reference term in the above equation is given by
\begin{equation}
G_h = F_h(N,V,T) + F_{\rm jel}[n_e,T] + PV,
\end{equation}
where $F_h(N,V,T)$ is the free energy of the harmonic lattice (Eq.~7 of Hamaguchi \textit{et al.} 1997 \cite{hamaguchi1997}) and the volume is found by numerically solving
\begin{equation}
- \left. \frac{\partial (F_h+F_{\rm jel})}{\partial V} \right|_{N,T} = P.
\label{eq:psolid}
\end{equation}
This time, the integration is performed numerically over the whole domain (i.e., from $\lambda=1$ to $\lambda \rightarrow \infty$, although in practice the integration can be stopped at $\lambda = 5$ where the anharmonic corrections become negligible). We use Simpson's rule and the results from 20 simulations from $\lambda=1$ to $\lambda = 5$, with an increased sampling for $1 \leq \lambda \leq 2$ where the integrand of Eq.~\eqref{eq:npt_int_solid} is more important.

\begin{figure}
\includegraphics[width=\columnwidth]{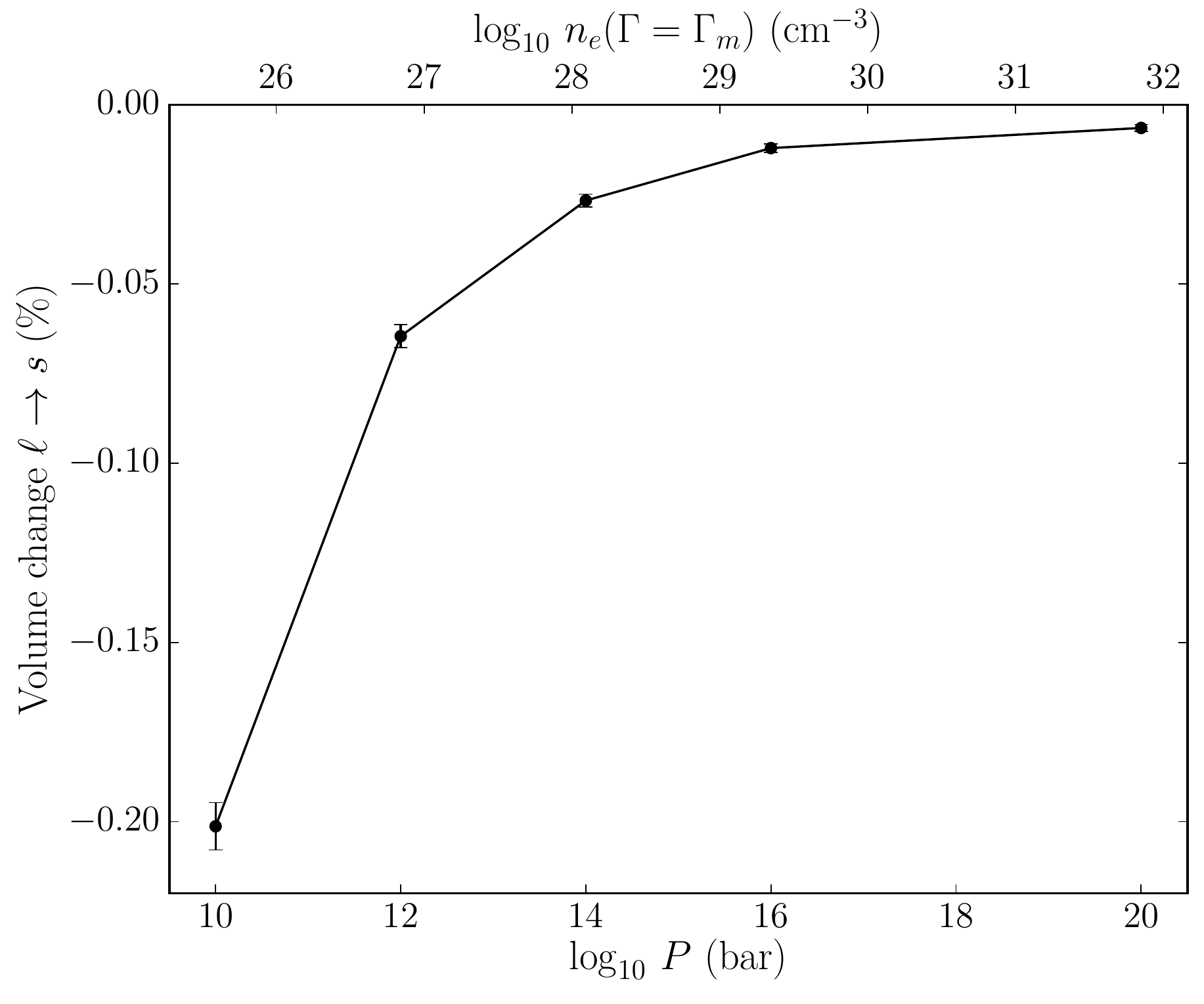}
\caption{Volume change of the OCP (with $Z=6$) during the liquid to solid phase transition at constant pressure. The lower horizontal axis gives the pressure imposed to the OCP and the upper axis gives the electron density at the temperature where the transition occurs.}
\label{fig:volume_npt}
\end{figure}

Unlike in the NVT case where only $\Gamma$ was changed from one simulation to the other, we must now vary $\lambda$, $T$, and $P$:
\begin{enumerate}
\item For a given $T$ and $P$, we perform simulations at different $\lambda$ values in order to obtain the Gibbs free energy $G(N,P,T)$ via integration of Eqs.~\eqref{eq:npt_integ_liq} and \eqref{eq:npt_int_solid};
\item Keeping $P$ constant, step~1 is performed again at a different $T$. This step is repeated until, for a given pressure, $G(T)$ can be satisfactorily mapped to find the temperature at which $G^{\ell}(T)$ and $G^{s}(T)$ intersect. 
\item Steps~1 and 2 are repeated for different pressures. This allows to find the melting temperatures at different pressures, thus enabling the study of the effect of volume changes on $\Gamma_m$.
\end{enumerate}
As for the NVT case, we found that using 1024 ions and 4000 MC cycles is more than enough to achieve convergence and sufficiently small statistical errors. All calculations presented below were obtained for a pure C plasma, but our conclusions remain unchanged if the ionic charge is changed. Fig.~\ref{fig:volume_npt} shows how the volume change across the phase transition varies as a function of pressure. As expected, the volume change is rather small and is smallest when the total pressure is very high and completely dominated by the electron gas. Our $\Delta V$ values are consistent with those obtained from a C equation of state used to model white dwarf interiors \citep{lamb1975}.

\begin{figure}
\includegraphics[width=\columnwidth]{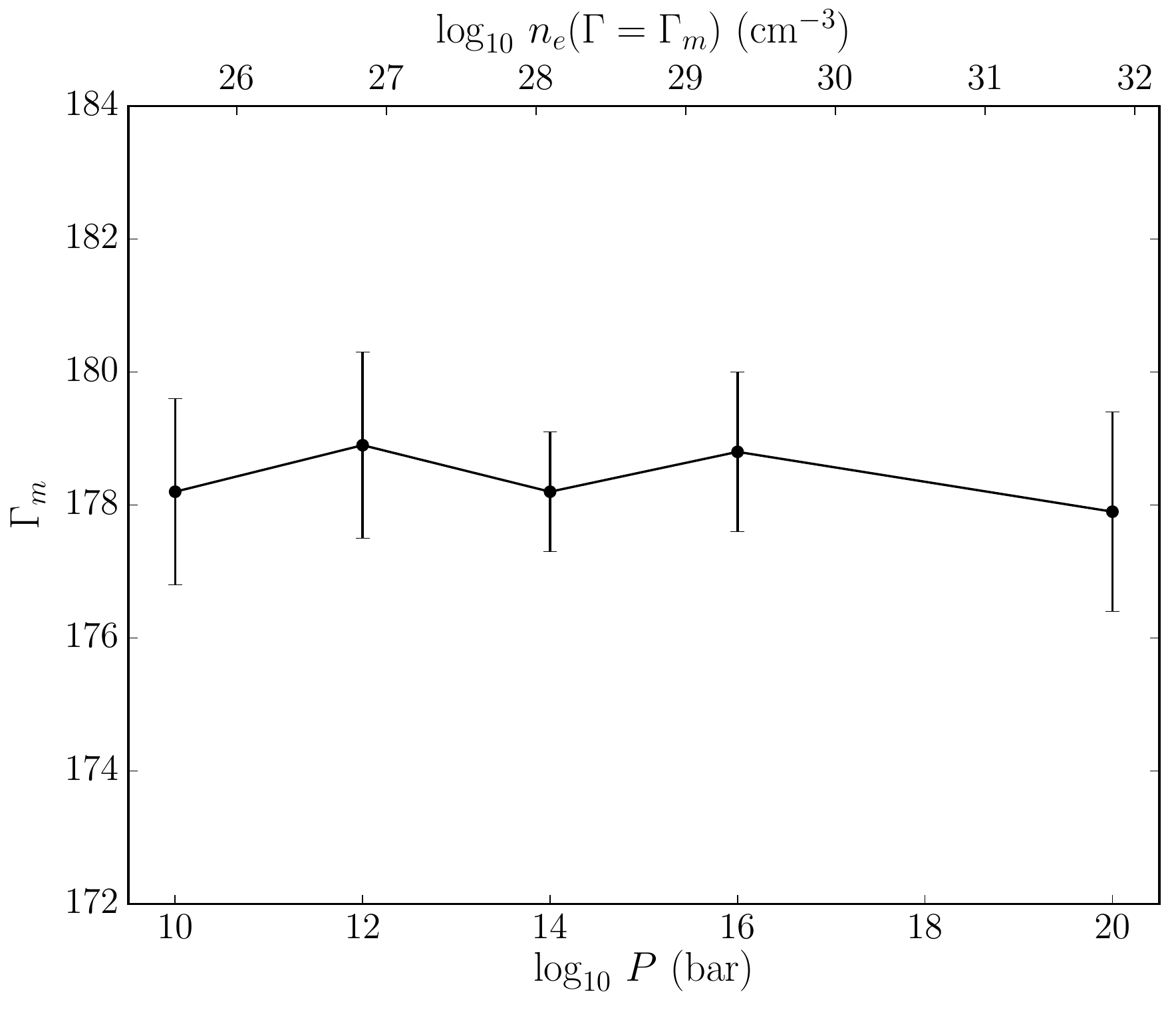}
\caption{Melting temperature of the OCP (with $Z=6$) as a function of pressure.}
\label{fig:gamma_npt}
\end{figure}

In Fig.~\ref{fig:gamma_npt}, we show the melting temperature of the C plasma as a function of pressure. Within a pressure range that spans 10 orders of magnitude, $\Gamma_m$ is constant within the statistical uncertainties: all our values are consistent with $\Gamma_m \simeq 178 \pm 2$. This value agrees with our NVT determination ($\Gamma_m = 174.6 \pm 1.6$) as well as with modern values published in the literature. Note that the small offset between our NVT and our NPT $\Gamma_m$ values is likely due to the many systematic differences between the two methods (different analytic fits, different thermodynamic integrations, and different ensembles). This numerical experiment clearly demonstrates that the volume change can indeed be safely ignored when modeling the OCP liquid--solid phase transition, which is consistent with the result of refs.~\cite{ogata1993,medin2010}. The term $P ( V^{\ell} - V^s )$ is by no means negligible, but $P ( V^{s} - V^{\ell} )\simeq -  (F_{\rm jel}^{s} - F_{\rm jel}^{\ell} )$, so that $G^{s} - G^{\ell} \simeq F^{s} - F^{\ell}$.

This exploration of the effect of the volume change on the melting temperature is somewhat artificial since we imposed $\kappa = 0$ for all pressures. While this test allowed us to isolate the effect of $\Delta V$, the OCP is not a good model at the lower end of the density range we studied (see Fig.~\ref{fig:kappa}). We have recalculated the melting curve of a C plasma with $\kappa$ dynamically adjusted during the NPT simulations to obey Eq.~\eqref{eq:kappa}. All the methodology presented above for the OCP remains the same, but we note that care must be taken when computing $\partial F_h / \partial V$ in Eq.~\eqref{eq:psolid} as $\kappa$ now depends on $V$.

\begin{figure}
\includegraphics[width=\columnwidth]{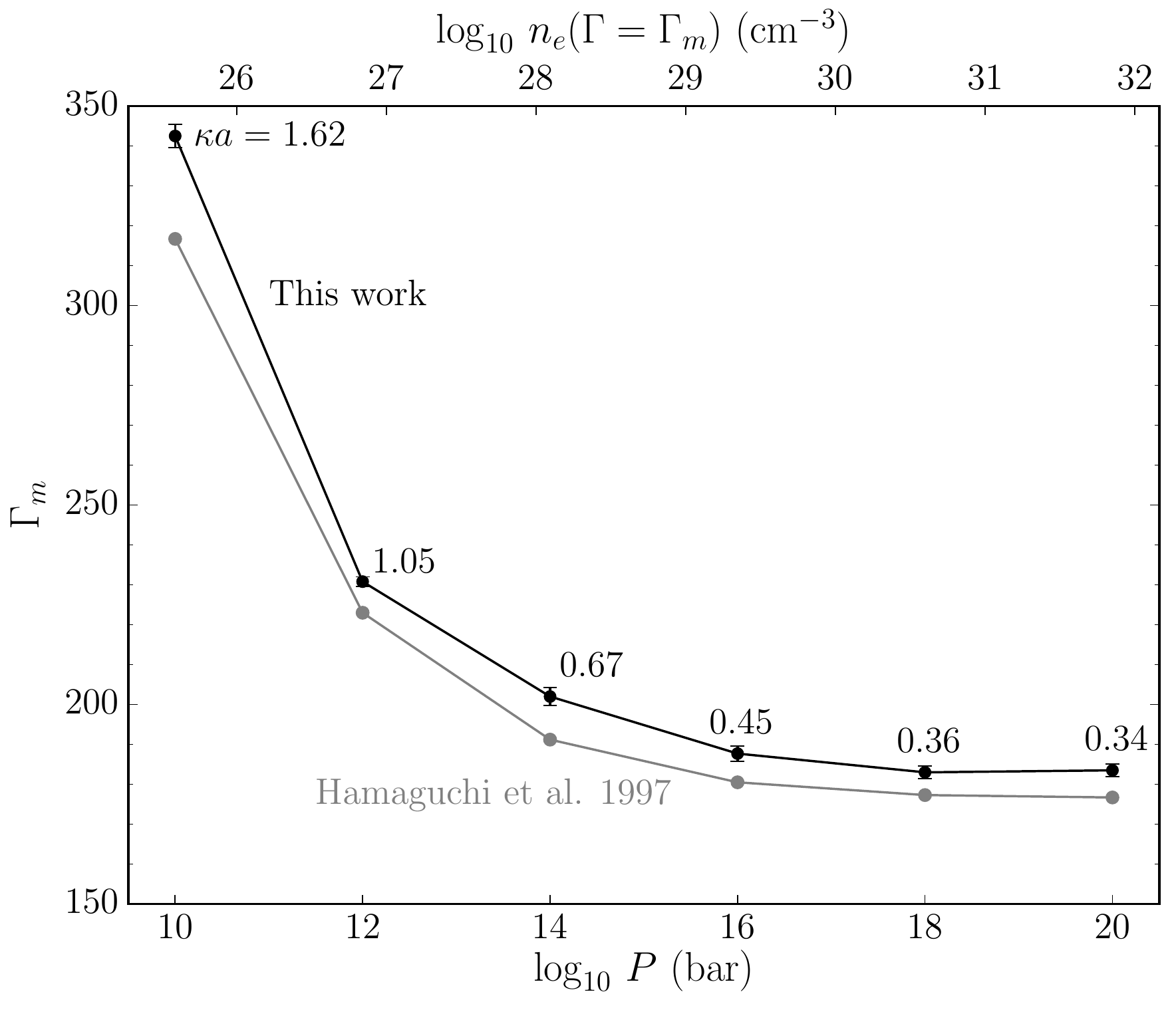}
\caption{Melting temperature of a Yukawa system (with $Z=6$) as a function of pressure. The screening length $1/\kappa$ is adjusted according to the electron density [Eq.~\eqref{eq:kappa}]. For comparison, the results of Hamaguchi \textit{et al.} (1997) \citep{hamaguchi1997} for the corresponding $\kappa a$ values are shown.}
\label{fig:gamma_npt_kappa}
\end{figure}

Once again, we find that volume changes can be neglected. Fig.~\ref{fig:gamma_npt_kappa} shows how the melting temperature changes as a function of pressure. The variation of $\Gamma_m$ is entirely due to the variation of $\kappa$ with pressure. For comparison, we show the results of Hamaguchi \textit{et al.} (1997) \citep{hamaguchi1997}, obtained from fits to MD simulations in the NVT ensemble. Both curves are very similar, although there is a small systematic offset in $\Gamma_m$ between our results and theirs. This difference is not surprising as our methods are completely different: (1) we use MC simulations while they use MD simulations, (2) we work in the NPT ensemble while they work in the NVT ensemble, and (3) we use different analytic fits and thermodynamic integrations.

\subsubsection{NPT$\Delta \mu$ Ensemble}
Finally, we turn to the validation of our simulations in the isobaric semi-grand canonical ensemble, the ensemble needed to use the Clapeyron integration method described in this work. Unfortunately, very few results of such simulations are published in the literature, but we were able to verify that our simulations are consistent with the results presented in Table~1 of ref.~\cite{kofke1988} for Lennard--Jones fluids.

\begin{figure}
\includegraphics[width=\columnwidth]{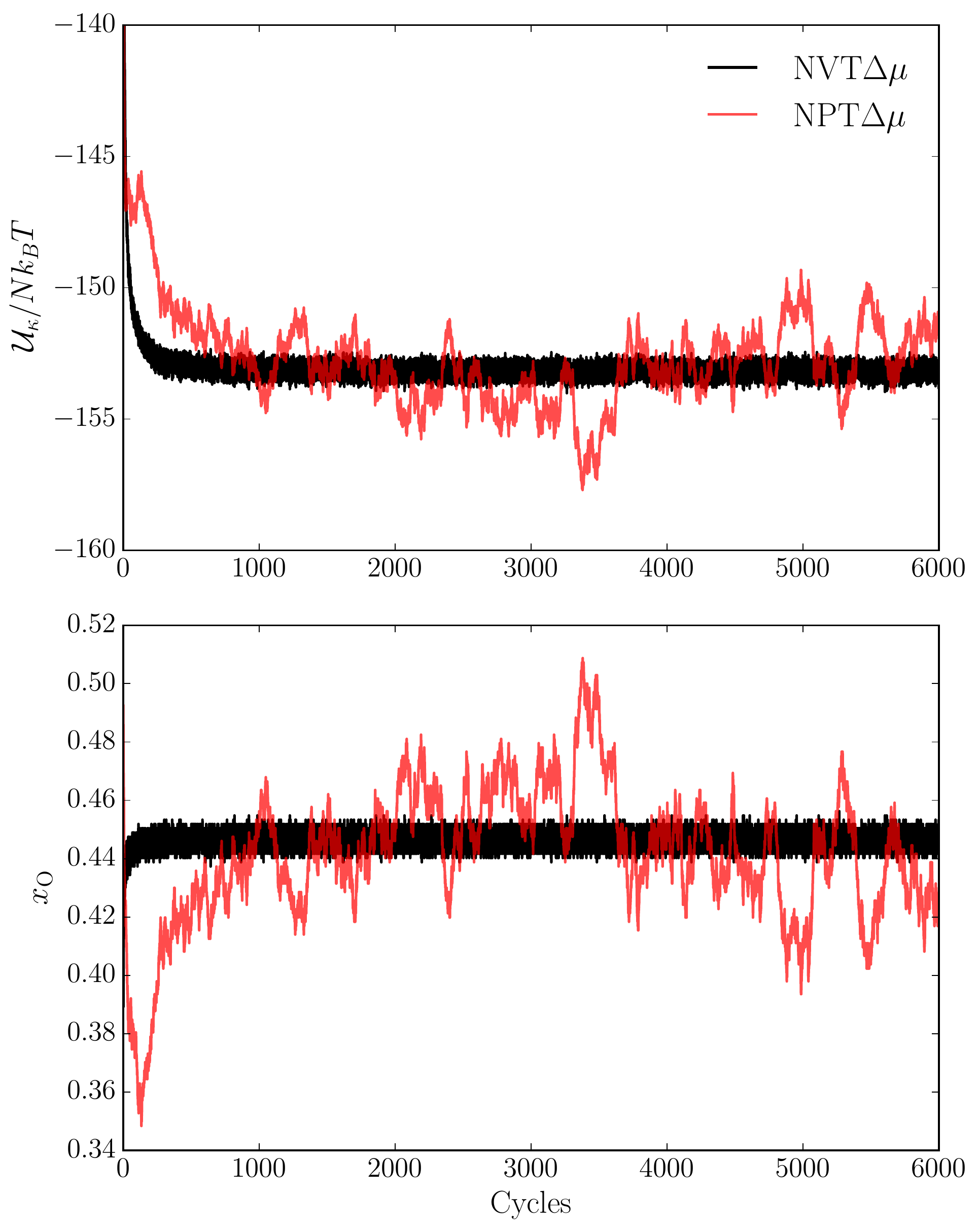}
\caption{Evolution of the ion excess energy and of the O concentration in isobaric (red) and isochoric (black) semi-grand canonical simulations of a C/O plasma. $N=686$, $T=409.4\,{\rm eV}$, $\xi_{\rm O}=0.5$, $\mu_e^{\rm ref}=523.7\,{\rm keV}$, $P=10^{18}\,{\rm bar}$ for the isobaric simulation and $N/V=4.967 \times 10^{29}\,{\rm cm}^{-3}$ for the isochoric simulation (which corresponds to the final density of the isobaric simulation).  ${\cal U}_{\kappa}/N k_B T=-153.169 \pm 0.004$ and $x_{\rm O}=0.44661 \pm 0.00002$ in the isochoric simulation, and ${\cal U}_{\kappa}/N k_B T=-153.1\pm0.3$ and $x_{\rm O}=0.446 \pm 0.005$ in the isobaric simulation.}
\label{fig:snpt_demo}
\end{figure}

To validate our simulations for electron--ion plasmas, we rely on comparisons to the somewhat simpler isochoric semi-grand canonical simulations (NVT$\Delta \mu$), where only particle displacements and identity changes are allowed. If, for a given temperature, total number of ions, and fugacity fraction, the volume imposed to the NVT$\Delta \mu$ simulation is the same as that obtained at the end of a NPT$\Delta \mu$, then the energies and concentrations of both simulations should agree. Fig.~\ref{fig:snpt_demo} demonstrates that it is indeed the case. Note that the convergence of the simulations is slower in the isobaric than in the isochoric case, because the parameter space to explore is larger in the former case as the volume is allowed to vary.

\clearpage

\bibliography{references}

\end{document}